\documentclass[a4paper]{article}
\usepackage[a4paper,top=3cm,bottom=2cm,left=3cm,right=3cm,marginparwidth=1.75cm]{geometry}
\usepackage{amsfonts}
\usepackage{bm}
\usepackage{extarrows}
\usepackage{amssymb}
\usepackage{color}
\usepackage[all]{xy}
\usepackage{graphicx}
\usepackage{braket}
\usepackage{amsmath}
\usepackage{appendix}
\usepackage{graphicx}
\usepackage[colorinlistoftodos]{todonotes}
\usepackage{amsthm}
\usepackage{mathrsfs}
\usepackage{amssymb}
\usepackage{accents}
\usepackage{bbm}
\usepackage{extarrows}
\usepackage{makecell}
\usepackage{authblk}
\usepackage{url}
\usepackage{graphicx}
\usepackage{hyperref}
\usepackage{cite}
\usepackage{eufrak}
\hypersetup{colorlinks,
            citecolor=black,
            linkcolor=black,
            urlcolor=green,
            pdftex}

\newtheorem{theorem}{Theorem}

\def\be{\begin{equation}}
\def\ee{\end{equation}}
\def\ba{\begin{eqnarray}}
\def\ea{\end{eqnarray}}

\usepackage[english]{babel}
\usepackage[utf8x]{inputenc}
\usepackage[T1]{fontenc}

\usepackage[normalem]{ulem}

\title{The semiclassical propagator for coherent state on twisted geometry}
\author[1]{Gaoping Long }
\author[2]{Hongguang Liu \footnote{hongguang.liu@gravity.fau.de}}
\author[2]{Cong Zhang \footnote{zhang.cong@mail.bnu.edu.cn}}

\affil[1]{College of Physics $\&$ Optoelectronic Engineering, Jinan University, Guangzhou, 510632, Guangdong, China}

\affil[2]{Department Physik, Institut f\"ur Quantengravitation, Theoretische Physik III, Friedrich-Alexander Universit\"at Erlangen-N\"urnberg, Staudtstr. 7/B2, 91058 Erlangen, Germany}
\date{}

\begin{document}

\maketitle

\begin{abstract}
A new set of twisted geometric variables is introduced to parametrize the holonomy-flux phase space in loop quantum gravity. It is verified that these new geometric variables, after symplectic reduction with respect to the Gauss constraint, form a Poisson algebra which is analogue to  that in quantum mechanics.  This property ensures that these new geometric variables provide a simple path measure, upon which  a new formulation of coherent state path integral based on  twisted geometry coherent state is established in loop quantum gravity. Especially, this path integral  is analytically computable by expanding the corresponding effective action around the complex evolution trajectories at second order, and the result gives the semi-classical approximation of the quantum propagator between twisted geometry coherent state in LQG.

\end{abstract}

\section{Introduction}
Loop quantum gravity (LQG) opens a convincing approach to achieve the unification of general relativity (GR) and quantum mechanics (QM) \cite{first30years,Ashtekar2012Background,RovelliBook2,Han2005FUNDAMENTAL,thiemann2007modern,rovelli2007quantum}. The distinguished feature of LQG is its non-perturbative and background-independent construction.
LQG begins by reformulating classical GR as a $SU(2)$ gauge theory, using the Ashtekar variables $(A, E)$. In this formulation, $A$ represents the $SU(2)$ connection, while $E$, which is conjugate to $A$, represents the densitized triad. Then, one chooses the holonomies of $A$ along all edges and the fluxes of $E$ across all 2-surfaces as the fundamental variables for canonical quantization. Crucially, these holonomies and fluxes are well-defined without requiring any extra background structures on the underlying manifold, ensuring the background-independent nature of the theory.
This theory provided several important breakthroughs (see, e.g. \cite{first30years} for review and see, e.g., \cite{Zhang:2021qul,Long:2021izw,Zhang:2022vsl,Liegener:2019jhj,Han:2024rqb,Han:2024ydv,Assanioussi:2015gka} for some recent developments).
For instance, a family of operators representing geometric observables (2-surface area, 3-region volume, inverse metric tensor) were regularized without need to subtract infinities and their spectra turned out to be discrete,  thereby predicting the fundamental discreteness of spatial geometry.\cite{Ashtekar:1996eg,Ashtekar:1997fb,Bianchi:2008es,Ma:2010fy,Giesel_2006Consistencycheck,Yang_2019Consistencycheck,ROVELLI1995593,QoperatorPhysRevD.62.104021,volumePhysRevD.94.044003,long2020operators}. 
Furthermore, LQG offers a microscopic description of spacetime thermodynamics. The corresponding state counting, based on boundary states, along with the computation of entanglement entropy from quantum geometry states in LQG, successfully reproduces black-hole entropy, including specific quantum corrections \cite{Ashtekar:1997yu,Ashtekar:2000eq, Long:2024lbd,Song:2020arr,Kaul:2000kf,Ghosh:2011fc,Basu:2009cw,Engle:2010kt,Song:2022zit,Donnelly:2008vx,Perez:2014ura,Dasgupta:2005yu}. 
Moreover, by applying the loop quantization method to symmetry-reduced models, such as in cosmology and black hole, LQG predicts intriguing phenomena like big bounce cosmology and black-to-white hole transitions \cite{Ashtekar:2003hd,Ashtekar:2011ni,Long:2020oma,Zhang:2021zfp,Bojowald:2001xe, Ashtekar:2006rx, Ashtekar:2006wn,Ashtekar:2005qt,Modesto:2005zm, Boehmer:2007ket,Chiou:2012pg,Gambini:2013hna,Corichi:2015xia,Dadhich:2015ora,Olmedo:2017lvt,Ashtekar:2018lag,BenAchour:2018khr,Han:2020uhb,Kelly:2020lec, Han:2022rsx,Giesel:2023hys,Ashtekar:2023cod,PhysRevLett.102.051301}.

The dynamics of the full theory of LQG face significant challenges due to the non-polynomial nature of the Hamiltonian constraint.
Recently, the coherent state path integral formulation has been applied to tackle this issue in relevant theories, i.e. the full theory, the cosmological models and the weak coupling model of LQG \cite{Han_2020,Han_2020semiclassical,Long:2021izw,Qin:2011hx,Qin:2012xh,Han:2021cwb,Han:2020iwk}. The coherent state path integral is derived within the context of the reduced phase space formulation of canonical LQG. 
 Specially, in the reduced phase space formulation, matter fields such as dust or scalar fields are introduced as clock fields, and the constraints are solved classically to obtain a physical Hamiltonian that governs the evolution with respect to the clock field. In the quantum theory, this physical Hamiltonian can be quantized, leading to a physical Hamiltonian operator. 
With this operator, the transition amplitude associated with its unitary evolution can be considered, where the initial and final states are given by coherent states. To compute this amplitude, the coherent state path integral formulation is then introduced.

 There are several proposals for constructing coherent states in LQG. The most widely used  coherent state in the (1+3)-dimensional LQG is the heat-kernel coherent state, which is first proposed by Hall and further applied to general gauge field theories by Thiemann  \cite{1994The,ThiemannComplexifierCoherentStates,Thomas2001Gauge,2001Gauge,2000Gauge}. It has been shown that the heat-kernel coherent state of $SU(2)$ possesses a well-behaved peakedness property in the holonomy-flux phase space, and the ``Ehrenfest property'' of this coherent state guarantee the coincidence between the expectation values of the operators  and their classical evaluations in the phase space \cite{Han_2020,Han_2020semiclassical,Zhang:2021qul}.  The previous work \cite{Han_2020} applied the  heat-kernel coherent state in the coherent state path integral formulation.
Then, in the resulting path integral formula, the stationary phase approximation is employed to find the ``classical path" that predominantly contributes to the amplitude. Explicitly, 
 the classical path is the solution to the semiclassical equations of motion (EOM) derived by vanishing the variation of the effective of the amplitude. 
  Similarly, one can also consider  the $U(1)^3$ heat-kernel coherent state  to construct the coherent state path integral, which gives the effective dynamics of the weak coupling theory of LQG \cite{Long:2021izw}.

 Apart from the heat-kernel coherent state, another type of coherent state in the (1+3)-dimensional LQG is introduced based on the twisted geometry parametrization of the holonomy-flux phase. Such type of coherent state is first introduced to analyze the asymptotics of the EPRL spin foam model \cite{Rovelli_2006,Bianchi_2009,Bianchi:2009ky}, and further studied comprehensively in \cite{Calcinari_2020}. This twisted geometry coherent state can be constructed by taking proper superposition of the spin-networks labelled by the coherent intertwiners at vertices \cite{Long:2020euh,long2019coherent,Calcinari_2020,Livine:2007Nsfv}, or equivalently, can be derived from the heat-kernel coherent state by selecting some special terms. The twisted geometry coherent state has a much simpler structure comparing to the heat-kernel coherent state, so that they can be directly extended to all dimensional LQG to avoid the complicatedness of the gauge group with higher rank \cite{PhysRevD.104.046014,Long:2021lmd,Long:2022cex}. Moreover, the completeness, peakedness and Ehrenfest properties of the twisted geometry coherent state are also established \cite{Calcinari_2020,Long:2021lmd,Long:2022cex}.

The twisted geometry coherent states provides us a new basement to construct the dynamics of LQG. In other words, we can construct the transition amplitude of the unitary evolution generated by the physical Hamiltonian operator, with the  initial and final states are given by the twisted geometry coherent states. Though such a formulation has been constructed for heat-kernel coherent state, the twisted geometry coherent state has its own advantages to ensure us make a further exploration in this direction. Specifically, the symplectic structure and the corresponding measure on the holonomy-flux phase space can be simplified by expressing them in terms of the twisted geometry variables. Further, one can separate the gauge degrees of freedom and physical degrees of freedom based on the twisted geometry, to establish a coherent state path integral formulation on the reduced phase space.
In this article,  we will first establish a set of new geometric variables to separate the gauge invariant geometry degrees of freedom and pure gauge degrees of freedom encoded in the twisted geometry. Especially, our results show that this set of new geometric variables forms a simpler  Poisson algebra and thus a simpler path measure in the reduced phase space with respect to Gauss constraint. Also, we will show that the overlap amplitude between twisted geometry coherent state reproduce the symplectic structure of the twisted geometry phase space. Thus, by following a similar procedure as in Ref.\cite{Han_2020,Long:2021izw}, we will construct the coherent state path integral by using the overlap amplitude and the over-completeness of twisted geometry coherent state, with an effective action being concluded from this coherent state path-integral formulation.  Further, notice that the Hamiltonian in the effective action is gauge invariant {under the transformation generated by } the Gauss constraint, {one can carry out the path integral over the gauge-variant sector directly, which leads to the reduced coherent state path integral define in the reduced phase space (with respect to the Gauss constraint).} Crucially, based on the path measure which is compatible to  the simpler Poisson algebra in the reduced phase space, we find that  the reduced path integral  is computable by expanding the corresponding effective action around the complex evolution trajectories at second order. By applying the relevant computation skills for coherent state path-integral to our case, we will get the semi-classical approximation of the quantum propagator between twisted geometry coherent state in LQG.

%Notice that the resulted effective action is a function of twisted geometry parameters and the corresponding resulted effective action given by heat-kernel coherent state is a function of holonomy-flux variables. Hence, this two effective actions can be compared based on the twisted geometry parametrization of holonomy-flux phase space. Our result shows that, the effective actions given by heat-kernel and twisted geometry coherent state path-integral respectively are consistent in continuum time limit up to higher order corrections of $t\propto\kappa\hbar$, provides that the expectation values of the Hamiltonian operator are well-evaluated by the corresponding classical value for both of these two types of coherent states.

This paper is organized as follows. In section \ref{sec:twoone}, we first will review the general structure of LQG, including the classical connection dynamics of GR and the basic elements in the loop quantized theory; Then, in the section \ref{sec:twotwo},  we will review the twisted geometry parametrization of the holonomy-flux phase space; Especially, in section \ref{sec:twothree}, we will introduce new geometric variables to parametrized the twisted geometry space. It will be verified that these new geometric variables, after symplectic reduction with respect to Gauss constraint, form a Poisson algebra which is analogue to  that in QM.
In section \ref{sec:three}, the over-completeness property  and the inner product of twisted geometry coherent state will be introduced; Especially, we will analyse the related phase space measure, symplectic structure and canonical transformations.  The path-integral based on twisted geometry coherent state will be established in section \ref{sec:four}; We will first construct the path-integral formulation in the twisted geometry phase space,  and then carry out the integral over the gauge variables to get the path-integral in the reduce phase space; Further, the equation of motions (EOMs) can be derived by varying  the effective action extracted from the reduced path integral, which give the complex evolution trajectories in the complexified reduced phase space; We will consider the second order expansion of the effective action around the  complex evolution trajectories, and the reduced path integral will be carried out based on this expansion, which lead to the semi-classical  propagator for twisted geometry coherent state.  Finally, we will finish with a summary and discussion in section \ref{sec:five}.

\section{Elements of LQG}
\subsection{The basic structures}\label{sec:twoone}

The (1+3)-dimensional Lorentzian LQG starts  with writing the Hamiltonian formulation of GR in terms of the Ashtekar variables $A_a^i$ and $E^b_j$, which are, respectively, the $SU(2)$ connection and densitized triad defined on the spatial manifold field $\Sigma$ \cite{Ashtekar2012Background,RovelliBook2,Han2005FUNDAMENTAL,thiemann2007modern,rovelli2007quantum}. They have the Poisson bracket
\begin{equation}
\{A_{a}^i(x),E^{b}_j(y)\}=\kappa\beta\delta_a^b\delta^i_j\delta^{(3)}(x-y).
\end{equation}
where $\kappa=8\pi G$ with $G$ being the Newton constant and $\beta$ is the Babero-Immirze parameter.
Here we use $i, j, k, ...$ for the internal $su(2)$ index and $a, b, c, ...$ for the spatial index. The Ashtekar new variables encode the 4-dimensional geometric, i.e., the 3-metric $q_{ab}$ and the extrinsic curvature $K_{ab}$ of $\Sigma$ as follows. At  first, the densitized triad  $E^a_i$ is related to the triad $e^a_i$, which is given by $q_{ab}=e_a^ie_{bi}$, by $E^{a}_{i}=\sqrt{\det(q)}e^{a}_{i}$, where $\det(q)$ denotes the determinant of $q_{ab}$. In addition, the connection can be expressed as $A_{a}^{i}=\Gamma_{a}^{i}+\beta K_{a}^{i}$, where $\Gamma_{a}^{i}$ is the spin connection compatible with $e_{a}^{i}$ and $K_a^i$ is defined by by $K_a^i=K_{ab}e^b_j\delta^{ji}$.
In this formulation, the dynamics of the gravitational fields is governed by the Gauss constraint $\mathcal {G}^i$, the vector constraint $\mathcal{C}_a$ and the scalar constraint $\mathcal{C}$, which read
\begin{equation}\label{GC}
 \mathcal{G}^i:=\partial_aE^{ai}+A_{aj}E^a_k\epsilon^{ijk}=0,
\end{equation}
\begin{equation}\label{VC}
 \mathcal{C}_a:=E^b_iF^i_{ab}=0,
\end{equation}
and
\begin{equation}\label{SC}
\mathcal{C}:=\frac{E^a_i E^b_j}{\det{(E)}}({\epsilon^{ij}}_kF^k_{ab}-2(1+\beta^2)K^i_{[a}K^j_{b]})=0,
\end{equation}
where $F_{ab}^i=\partial_aA_b^i-\partial_bA_a^i+\epsilon_{ijk}A_a^jA_b^k$ is the curvature of $A_a^i$.
As it is totally constrained, the system suffers the problem of time \cite{thiemann2007modern}. This issue can be addressed by introducing extra matter fields, which serve as physical time for deparametrization \cite{Brown:1994py,PhysRevD.43.419,Domagala:2010bm}. In the  deparametrized models, the physical Hamiltonian $\mathbf{H}$, governing the evolution with respect to the physical time, takes the form $\mathbf{H}=\int_{\Sigma}dx^3h$, where the densitized scalar field $h=h(\mathcal{C},\mathcal{C}_a)$ takes different formulations for different deparametrization models. For instance, in the model deparametrized by the Gaussian dust, one has $h=\mathcal{C}$ \cite{PhysRevD.43.419,Han_2020}.

 Performing the loop quantization for the phase space of the Ashtekar new variables, we get the kinematic Hilbert space of LQG $\mathcal H$. The states in the Hilbert space are defined based on graphs embedded in $\Sigma$. The states based on a certain graph $\gamma$ form the Hilbert subspace $\mathcal H_\gamma=L^2(SU(2)^{|E(\gamma)|},d\mu_{\text{Haar}})$, where $E(\gamma)$ denotes the set  of elementary edges of $\gamma$ so that $|E(\gamma)|$ is the number of elements in $E(\gamma)$, and $d\mu_{\text{Haar}}$ is the  the Haar measure on $SU(2)$. In the definition of the Hilbert space $\mathcal H_\gamma$, we distribute to each edge of the graph $\gamma$ an manifold of $SU(2)$ as the configuration space associated with the edge.
In this sense, on each given $\gamma$ there is a discrete phase space $(T^\ast SU(2))^{|E(\gamma)|}$, which is coordinatized by the basic discrete variables---holonomies and fluxes. Specifically, the holonomy of $A_a^i$ along an edge $e\in\gamma$ is defined by
 \begin{equation}
h_e[A]:=\mathcal{P}\exp(\int_eA)=1+\sum_{n=1}^{\infty}\int_{0}^1dt_n\int_0^{t_n}dt_{n-1}...\int_0^{t_2} dt_1A(t_1)...A(t_n),
 \end{equation}
 where $A(t)=A_a^i(t)\dot{e}^a(t)\tau_i$, and $\tau_i=-\frac{\mathbf{i}}{2}\sigma_i$ with $\sigma_i$ being the Pauli matrices. The covariant flux associated with an edge $e$ can be defined in two ways \cite{Thomas2001Gauge,Zhang:2021qul}.  The first one, denoted by $F^i_e$  or ${ F}^i(e)$, is defined to be associated with the source point as
 \begin{equation}\label{F111}
 F^i_e:=\frac{2}{\beta }\text{tr}\left(\tau^i\int_{S_e}\epsilon_{abc}h(\rho^s_e(\sigma))E^{cj}(\sigma)\tau_jh(\rho^s_e(\sigma)^{-1})\right),
 \end{equation}
 where $S_e$ is the 2-face in the dual lattice $\gamma^\ast$ of $\gamma$, $\rho^s(\sigma): [0,1]\rightarrow \Sigma$ is a path connecting the source point $s_e\in e$ to $\sigma\in S_e$ such that $\rho_e^s(\sigma): [0,\frac{1}{2}]\rightarrow e$ and $\rho_e^s(\sigma): [\frac{1}{2}, 1]\rightarrow S_e$.  Similarly,   the second one, denoted by $\tilde{F}^i_e$  or $\tilde{F}^i(e)$, is defined to be associated with the target point as
  \begin{equation}\label{F222}
 \tilde{F}^i_e:=-\frac{2}{\beta }\text{tr}\left(\tau^i\int_{S_e}\epsilon_{abc}h(\rho^t_e(\sigma))E^{cj}(\sigma)\tau_jh(\rho^t_e(\sigma)^{-1})\right),
 \end{equation}
 where $\rho^t(\sigma): [0,1]\rightarrow \Sigma$ is a path connecting the target point $t_e\in e$ to $\sigma\in S_e$ such that $\rho_e^t(\sigma): [0,\frac{1}{2}]\rightarrow e$ and $\rho_e^t(\sigma): [\frac{1}{2}, 1]\rightarrow S_e$. It is easy to see that  these two fluxes can be related by
 \begin{equation}\label{F333}
 \tilde{F}^i_e\tau_i=-h_e^{-1} {F}^i_e\tau_ih_e.
 \end{equation}
 The non-vanishing Poisson brackets between the holonomy and fluxes read
 \begin{eqnarray}\label{hp1220}
&&\{{{h}}_e[{A}],{{F}}^i_{e'}\}= -\delta_{e,e'}{\kappa}\tau^i{{h}}_e[{A}],\quad \{{{h}}_e[{A}],\tilde{{F}}^i_{e'}\}= \delta_{e,e'}{\kappa}{{h}}_e[{A}]\tau^i,\\\nonumber
&&\{{{F}}^i_{e},{{F}}^j_{e'}\}= -\delta_{e,e'}{\kappa}{\epsilon^{ij}}_k{{F}}^k_{e'},\quad \{\tilde{{F}}^i_{e},\tilde{{F}}^j_{e'}\}= -\delta_{e,e'}{\kappa}{\epsilon^{ij}}_k\tilde{{F}}^k_{e'}.
\end{eqnarray}

\subsection{Twisted geometric parametrization of $SU(2)$ holonomy-flux phase space}\label{sec:twotwo}
For a fixed graph $\gamma$ , one has the  associated holonomy-flux phase space, which is coordinatized by the classical holonomy and flux. The holonomy-flux variables catch the discrete geometry information of the dual lattice of $\gamma$, which can be explained by the so-called twisted geometry parametrization of the holonomy-flux phase space \cite{PhysRevD.82.084040,PhysRevD.103.086016}.
Let us give a brief introduction of this parametrization as follow.

From now on, let us focus on a graph $\gamma$ whose dual lattice gives a partition of $\Sigma$ constituted by 3-dimensional polytopes. Elementary edge are referred to as these passing through exactly one 2-dimensional face in the dual lattice of $\gamma$. The discrete phase space related to the given graph $\gamma$ is  $\times_{e\in E(\gamma)}T^\ast SU(2)_e$ with $e$ being the elementary edges of $\gamma$. The symplectic potential on the phase space reads
\begin{equation}\label{sym1}
  \Theta_{\gamma}=\frac{a^2}{\kappa}\sum_{e\in E(\gamma)}\text{Tr}(p_e^i\tau_idh_eh_e^{-1}),
\end{equation}
where $p_e^i:=\frac{F_e^i}{a^2}$ is the dimensionless
flux with $a$ being a constant with the dimension of length, and $\text{Tr}(XY):=-2\text{tr}_{1/2}(XY)$ with $X,Y\in su(2)$.
Without loss of generality, we can first focus on the space $T^\ast SU(2)_e$ related to one single elementary edge $e\in E(\gamma)$.
This space can be parametrized by using the so-called twisted geometry variables as
 \begin{equation}
 (V_e,\tilde{V}_e,\xi_e, \eta_e)\in P_e:=S^2_e\times S^2_e\times T^\ast S^1_e,
 \end{equation}
 where $\eta_e\in\mathbb{R}$,  $\xi_e\in [0,2\pi)$, and
  \begin{equation}
  V_e:=V_e^i\tau_i, \   \ \  \tilde{V}_e:=\tilde{V}^i_e\tau_i,
  \end{equation}
 with $V_e^i,\tilde{V}^i_e\in S^2_e$ being unit vectors. 
 To relate the holonomy-flux variables with the twisted geometry variables, we need  to specify two sections $n_e:S^2_e\to SU(2)$ and $\tilde{n}_e:S^2_e\to SU(2)$ such that $V^i_e\tau_i=n_e(V_e)\tau_3n_e(V_e)^{-1}$ and $\tilde{V}^i_e\tau_i=-\tilde{n}_e(\tilde{V}_e)\tau_3\tilde{n}_e(\tilde{V}_e)^{-1}$. Then, the twist geometry variables can be related to the holonomy-flux by the map

\begin{eqnarray}\label{para}
(V_e,\tilde{V}_e,\xi_e,\eta_e)\mapsto(h_e,p^i_e)\in T^\ast SU(2)_e:&& p^i_e\tau_i=\eta_e V_e=\eta_en_e(V_e)\tau_3n_e(V_e)^{-1}\\\nonumber
&&h_e=n_e(V_e)e^{\xi_e\tau_3}\tilde{n}_e(\tilde{V}_e)^{-1}.
\end{eqnarray}
Note that this map is two-to-one map. In other words, \eqref{para} will map two points $\pm (V_e,\tilde{V}_e,\xi_e,\eta_e)\in P_e$ to the same point $(h_e,p^i_e)\in T^\ast SU(2)_e$.
Hence, by selecting either branch between the two signs related by a $\mathbb{Z}_2$ symmetry, one can establish a bijection  from the region in $P_e$ with either $\eta_e>0$ or $\eta_e<0$ to  $T^*SU(2)_e$ \cite{PhysRevD.82.084040,PhysRevD.103.086016}.

 Now we can get back to the discrete phase space of  LQG on the whole graph $\gamma$. It is just the Cartesian product of the discrete phase space on each single elementary  edge of $\gamma$. Namely, the discrete phase space on $\gamma$ can be parametrized by the twisted geometry variables in $P_\gamma:=\times_{e\in E(\gamma)}P_e$. To see how the variables in  $P_\gamma$ describe the geometry of the lattice dual $\gamma$, we note that  $|\eta_e|$ is interpreted as the dimensionless area of the 2-dimensional face dual to $e$, leading to the interpretation of $\eta _e V_e$ and $\eta _e \tilde{V}_e$ as the area-weighted outward normal vectors of the face in the frames of source and target points of $e$ respectively.
 Then, the holonomy $h_e=n_e(V_e)e^{\xi_e\tau_3}\tilde{n}^{-1}_e(\tilde{V}_e)$ rotates the inward normal $-\eta _e\tilde{V}_e$ to the outward normal $\eta _e{V}_e$ , i.e.,
  \begin{equation}
  \tilde{V}_e=-h_e^{-1}V_eh_e.
  \end{equation} 
   The symplectic 1-form on $P_\gamma$ resulting from \eqref{sym1} and the identification \eqref{para} reads
  \begin{equation}\label{sym2}
  \Theta_{P_\gamma}=\frac{a^2}{\kappa}\sum_{e\in E(\gamma)}\left(\eta_e\text{Tr}(V_edn_en_e^{-1})+\eta_ed\xi_e +\eta_e\text{Tr}(\tilde{V}_ed\tilde{n}_e\tilde{n}_e^{-1})\right),
\end{equation}
where $\text{Tr}(V_edn_en_e^{-1})$ and $\text{Tr}(\tilde{V}_ed\tilde{n}_e\tilde{n}_e^{-1})$ are the standard symplectic 1-form on the unit sphere $S^2_e$ \cite{PhysRevD.82.084040,PhysRevD.103.086016}.  In what following, the branch of $P_e$ with $\eta_e>0$ will be identified with $P_e^+:=(S^2_e\times S^2_e\times \mathbb{R}^+_{e}\times S^1_e)$.

It is worth to considering the gauge reduction with respect to the Gauss constraint in the twisted geometry space. The discrete Gauss constraint in the holonomy-flux phase
space $\times_{e\in E(\gamma)}T^\ast SU(2)_e$ is given by
 \begin{equation}
G_v:= -\sum_{e,s(e)=v}{p}^i_e\tau_i+\sum_{e,t(e)=v}p^i_eh_e^{-1}\tau_ih_e=0.
 \end{equation}
 The gauge transformation $\{g_v|v\in V(\gamma)\}$ generated by the Gauss constraint is
 \begin{equation}
 h_e\mapsto g_{s(e)}h_eg_{t(e)}^{-1},\quad {p}^i_e\tau_i\mapsto {p}^i_e g_{s(e)}\tau_i g_{s(e)}^{-1}.
 \end{equation}
 Correspondingly, {the gauge transformation will transform the twisted geometry variables as}
 \begin{eqnarray}\label{gaugetransV}
 % \nonumber to remove numbering (before each equation)
  {V}^i_e\tau_i\mapsto V_e(g_{s(e)}):= {V}^i_e g_{s(e)}\tau_i g_{s(e)}^{-1},  &&  \tilde{V}^i_e\tau_i\mapsto \tilde{V}_e(g_{t(e)}):=\tilde{V}^i_e g_{t(e)}\tau_i g_{t(e)}^{-1},\\\nonumber
  \xi_e\mapsto   \xi_e+\xi_{g_{s(e)}}-\xi_{g_{t(e)}}, &&  \eta_e\mapsto\eta_e,
 \end{eqnarray}
 where $\xi_{g_{s(e)}}$ and $\xi_{g_{t(e)}}$ are determined respectively by
 \begin{eqnarray}\label{xige}
 % \nonumber to remove numbering (before each equation)
 && g_{s(e)}n_e(V_e)=n_e(V_e(g_{s(e)}))e^{\xi_{g_{s(e)}}\tau_3},\\\nonumber
  &&g_{t(e)}\tilde{n}_e(\tilde{V}_e)=\tilde{n}_e(\tilde{V}_e(g_{t(e)}))e^{\xi_{g_{t(e)}}\tau_3}.
 \end{eqnarray}
 Then,  one can carry out the gauge reduction   {for }the discrete Gauss constraint {in $P_\gamma^+:=\times_{e\in E(\gamma)}P_e^+$}, which leads to the reduced phase space
 \begin{equation}
{H}^+_\gamma:={P}^+_\gamma/\!/SU(2)^{|V(\gamma)|}=\left(\times_{e\in E(\gamma)} T^\ast S_e^1\right)^+\times \left(\times_{v\in V(\gamma)} \mathfrak{P}_{\vec{\eta}_v}\right)
\end{equation}
 { where} $V(\gamma)$ is the collection of vertices in $\gamma$,  $|V(\gamma)|$ is the number of the vertices in $\gamma$ and
 \begin{equation}
 \mathfrak{P}_{\vec{\eta}_v}:=\{(V_{e_1},...,V_{e_{n_v}})\in \times_{e\in\{e_v\}}S^{2}_{e}| G_{v}=0 \}/SU(2).
 \end{equation}
{Here}  we re-oriented the edges linked to $v$ to be out-going at $v$ without loss of generality and $\{e_v\}$ represents the set of edges beginning at $v$ with $n_v$ being the number of elements in $\{e_v\}$.

\subsection{Reduced twisted geometry variables }\label{sec:twothree}
We still need to find the gauge invariant twisted geometry variables to parametrize the reduced phase space ${H}^+_\gamma$.  { To begin with, we need to factor out the holonomy $h_e^{\Gamma}$ of the spin connection $\Gamma$ from that of the Ashtekar connection $A=\Gamma+\gamma K$} \cite{Rovelli:2010km,PhysRevD.87.024038}. {To achieve this,} let us focus on a single edge $e$ and choose a gauge at $s(e)$ and $t(e)$ to ensure $V_e=-\tilde{V}_e$, {so that} the holonomy $h_e$ takes  $$h_e=n_e(V_e)e^{\xi_e\tau_3}n_e(V_e)^{-1}=e^{\xi_eV_e}$$ in this gauge.  {Also, in this gauge,} the contribution of extrinsic curvature to the holonomy $h_e$   is $e^{\varsigma_eV_e}$  \cite{Rovelli:2010km}. Consequently, the holonomy $h_e^{\Gamma}$ of the spin connection in this gauge is given by $$h_e^{\Gamma}=e^{\zeta_e V_e}\equiv e^{(\xi_e-\varsigma_e) V_e}.$$ {Finally, performing the gauge transformation as shown in \eqref{gaugetransV} to release the gauge aforementioned to ensure $V_e=-\tilde{V}_e$}, one can immediately get the general formulation for the holonomy $h_e^{\Gamma}$ of the spin connection in twisted geometry, which reads
\begin{equation}\label{hnn}
h_e^{\Gamma}=n_e(V_e)e^{\zeta_e \tau_3}\tilde{n}_e(\tilde{V}_e)^{-1}.
\end{equation}
Moreover,  {according to the gauge transformation for $\{(\xi_e,V_e, \tilde{V}_e)|e\in E(\gamma)\}$ given by Eqs. \eqref{gaugetransV}},  the holonomy $h^{\Gamma}$ of spin connection is transformed as
 \begin{equation}\label{gaugetranshgamma}
 h^{\Gamma}_e\mapsto g_{s(e)}h^{\Gamma}_eg_{t(e)}^{-1}.
 \end{equation}
Correspondingly,
 $\zeta_e$ is transformed as
\begin{equation}\label{zetatrans}
\zeta_e\mapsto   \zeta_e+\xi_{g_{s(e)}}-\xi_{g_{t(e)}}
\end{equation}
with $\xi_{g_{s(e)}}$ and $\xi_{g_{t(e)}}$ being given by Eq. \eqref{xige}.
Now, it is worth to giving the specific definition of $\zeta_e$ in $h_e^{\Gamma}$ for the cubic graph on the manifold with topology $\mathbb{T}^3$. Since the gauge transformation of $\zeta_e$ is known, we just need to define $\zeta_e$ in a specific gauge.  Let us still choose the gauge at $s(e)$ and $t(e)$ which ensures $V_e=-\tilde{V}_e$; Then, by choosing  a minimal loop $\square_e$ containing $e$, $e_s$ and $e_t$ with $s(e_s)=s(e)$ and $s(e_t)=t(e)$, the angle $\zeta_e=\zeta_{e,\square_e}\in(-\pi,\pi]$ can be determined by 
\begin{equation}
\cos(\zeta_e)=\frac{\delta^{ii'}\epsilon_{ijk}V^j_{e_s}V^k_e\epsilon_{i'j'k'}\tilde{V}^{j'}_{e}V^{k'}_{e_t}}{|\epsilon_{ijk}V^j_{e_s}V^k_e|\cdot|\epsilon_{i'j'k'}\tilde{V}^{j'}_{e}V^{k'}_{e_t}|}
\end{equation}
and 
\begin{equation}
\text{sgn}(\zeta_e)=\text{sgn}(\epsilon_{ijk}V_{e_s}^iV_{e_t}^jV_e^k)
\end{equation}
for the gauge aforementioned
which ensures  $V_e=-\tilde{V}_e$.  The geometric interpretation of  $\zeta_e$ is explained in Appendix \ref{app1}, where one can verify that the gauge transformation  \eqref{zetatrans} of  $\zeta_e$  is adapted to its geometric interpretation.

\begin{figure}[h]
 \centering
 \includegraphics[scale=0.15]{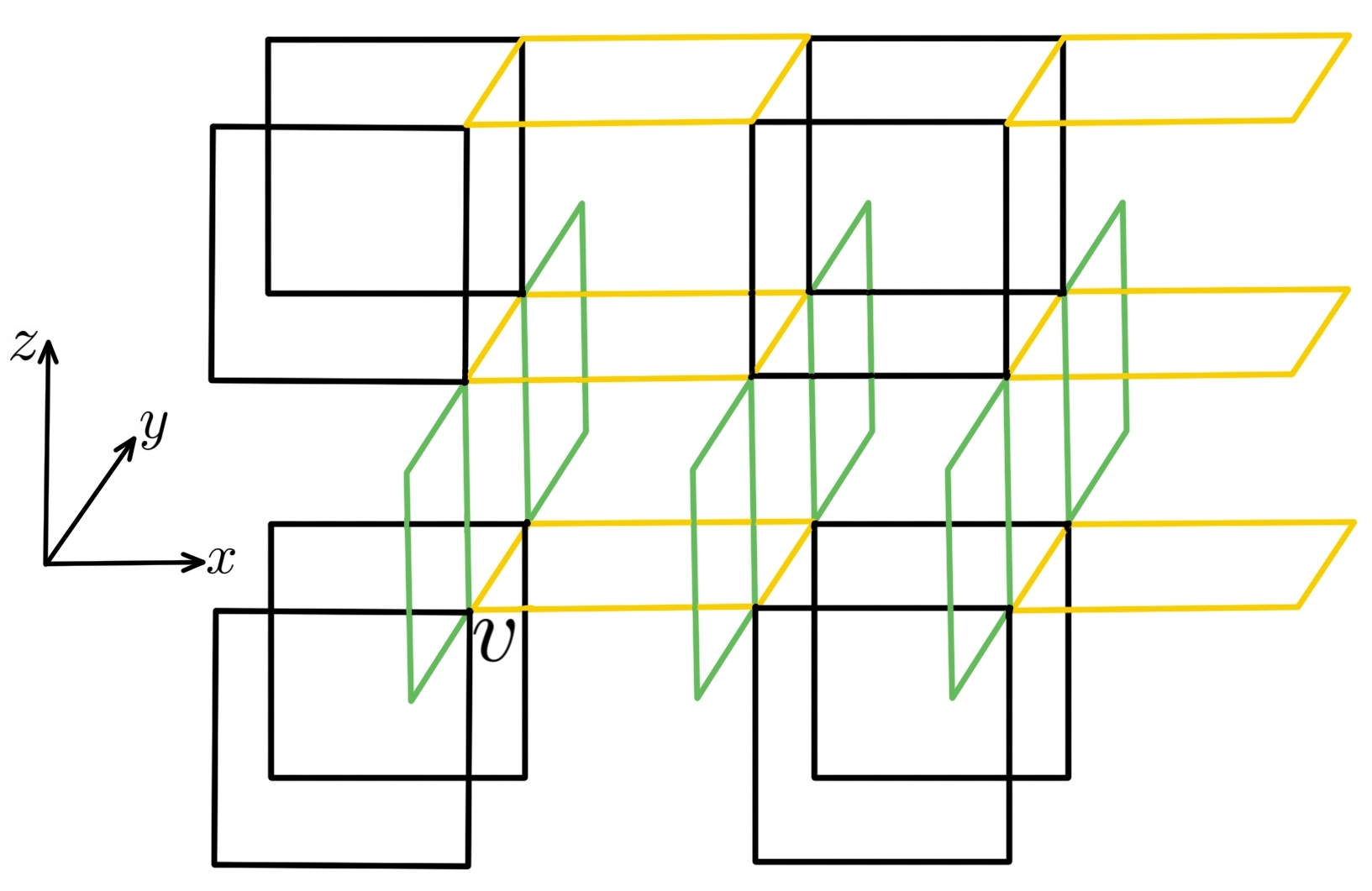}
\caption{The cubic graph can be regarded as a composition of square loops, with each edge belonging and only belonging to a square loop. For instance, the 6 edges linked to $v$ are assigned to 3 loops, which are marked with green, yellow and black respectively.}
\label{fig:label1}
\end{figure}

{One should notice that the definition of $\zeta_e=\zeta_{e,\square_e}$ depends on a choice of the minimal loop $\square_e$ containing $e$. Especially,  this choice can be fixed for the cubic graph on the manifold with topology $\mathbb{T}^3$.} The cubic   graph can be adopted to a Cartesian coordinate $\{x,y,z\}$, and then the edges linked to $v$ in this graph can be denoted by $e(v,\pm x)$, $e(v,\pm y)$ and $e(v,\pm z)$. More explicitly,  $e(v,+x)$ represents the elementary edge started at $v$ along $x$ direction,  $e(v,-x)$ represents the elementary edge ended at $v$ along $x$ direction, and likewise for $e(v,\pm y)$ and $e(v,\pm z)$. Now, we can fix the choice of the the minimal loop $\square_e$ in the definition of  $\zeta_e=\zeta_{e,\square_e}$ for every edge in the cubic graph, as shown in Fig.\ref{fig:label1} .
For instance,  the edges linked to the vertex $v$ in Fig.\ref{fig:label1}  has the following minimal loops
  \begin{equation}
   \square_{e(v,+y)}=\square_{e(v,+x)}\supset \{e(v,+x), e(v,+y)^{-1}\},
\end{equation}
 \begin{equation}
  \square_{e(v,-z)}=\square_{e(v,-x)}\supset \{e(v,-x), e(v,-z)^{-1}\},
\end{equation}
and 
 \begin{equation}
\square_{e(v,+z)}=  \square_{e(v,-y)}\supset \{e(v,-y), e(v,+z)\}.
\end{equation}
To simplify the notations, let us  denote $e_1(v)=e(v,+x)$, $e_2(v)=e(v,+y)$, $e_3(v)=e(v,-x)$, $e_4(v)=e(v,-z)$, $e_5(v)=e(v,-y)$, $e_6(v)=e(v,+z)$. Then, we can introduce the new notation  $\mathcal{V}^i_{e(v)}\equiv {V}^i_{e(v)}$ for $s(e)=v$ and $\mathcal{V}^i_{e(v)}\equiv \tilde{V}^i_{e(v)}$ for $t(e)=v$ .  

\begin{figure}[h]
 \centering
 \includegraphics[scale=0.25]{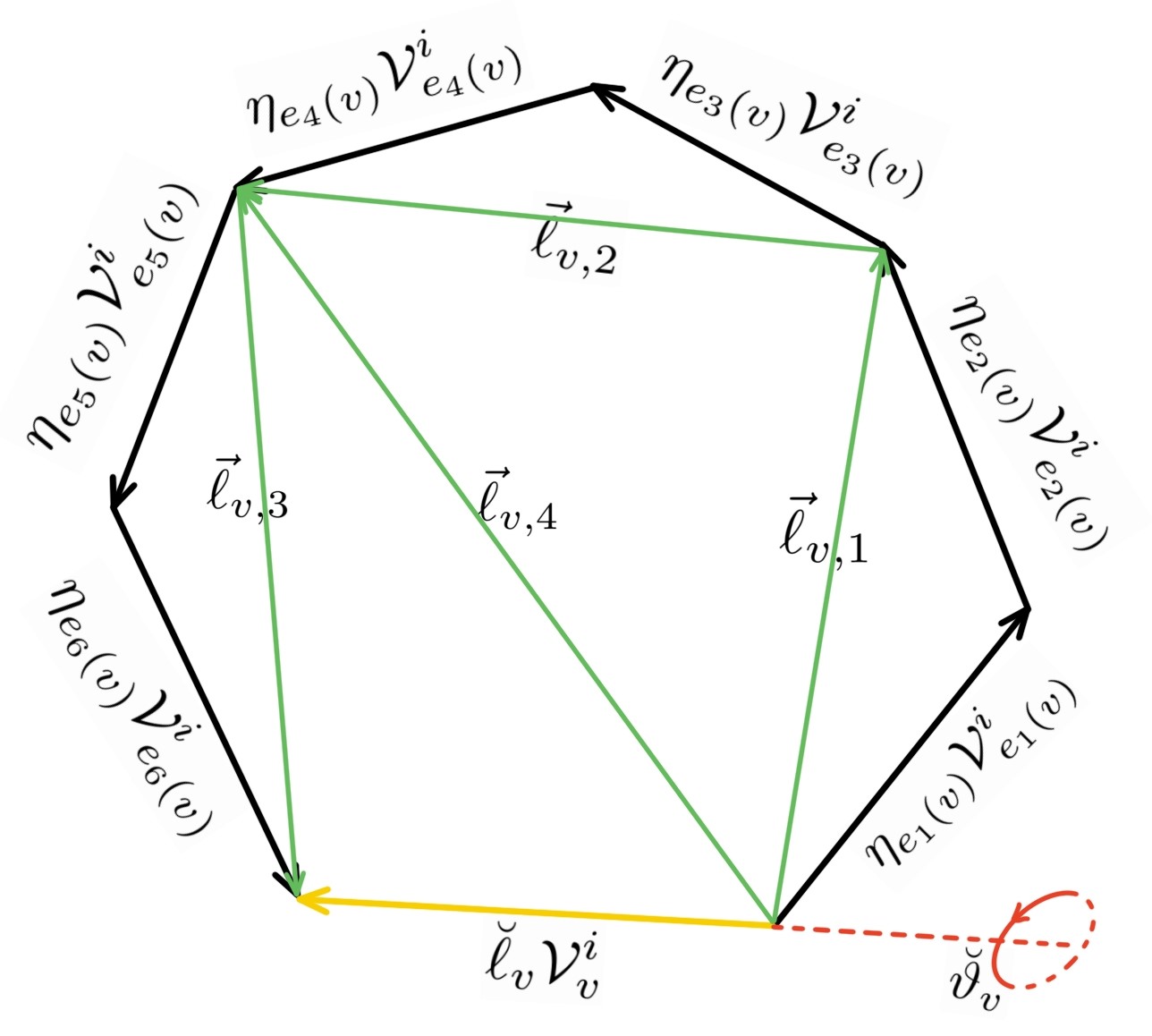}
\caption{The twisted geometry associated to the vertex $v$.}
\label{fig:label2}
\end{figure}

Now, let us start to introduce the reduced twisted geometry variables which parametrize $H_\gamma^+$. We first recall
\begin{eqnarray}
\label{decomp3}
\varsigma_e=\xi_e- \zeta_e.
\end{eqnarray}
{Due to $h_e= h^{\Gamma}_{e}e^{-(\xi_e- \zeta_e) \tilde{V}_e} = e^{(\xi_e- \zeta_e) {V}_e}h^{\Gamma}_{e}$, it can be checked that}
  \begin{equation}\label{repoi1}
 \{\varsigma_e,\eta_{e'}\}=\delta_{e,e'} \frac{\kappa}{a^2},\quad {\{\varsigma_e,G_v\}=0}.
\end{equation}
Then, we introduce the action-angle variables to re-parametrize the unit normal vectors $(\mathcal{V}^i_{e_1(v)},...,\mathcal{V}^i_{e_6(v)})$ as follows. 
Let us define 
  \begin{eqnarray}
 \vec{ \ell}_{v,1}\equiv  { \ell}^i_{v,1}&:=&\eta_{e_1(v)}\mathcal{V}^i_{e_1(v)}+\eta_{e_2(v)}\mathcal{V}^i_{e_2(v)},
\end{eqnarray}
 \begin{eqnarray}
 \vec{ \ell}_{v,2}\equiv  { \ell}^i_{v,2}&:=&\eta_{e_3(v)}\mathcal{V}^i_{e_3(v)}+\eta_{e_4(v)}\mathcal{V}^i_{e_4(v)},
\end{eqnarray}
 \begin{eqnarray}
 \vec{ \ell}_{v,3}\equiv  { \ell}^i_{v,3}&:=&\eta_{e_5(v)}\mathcal{V}^i_{e_5(v)}+\eta_{e_6(v)}\mathcal{V}^i_{e_6(v)},
\end{eqnarray}
 \begin{eqnarray}
 \vec{ \ell}_{v,4}\equiv  { \ell}^i_{v,4}&:=&{ \ell}^i_{v,1}+{ \ell}^i_{v,2},
\end{eqnarray}
and 
\begin{equation}
{\ell}_{v,I}=\sqrt{\delta_{ij}{\ell}_{v,I}^i{\ell}_{v,I}^j},\ \text{for}\ I=1,2,3,4.
\end{equation}
In addition, as shown in Fig.\ref{fig:label2}, we can define \cite{Bianchi:2010gc}:
\begin{itemize}
    \item[(1)] $\vartheta_{v,1}$  as the angle between the plane identified by the vectors $ (\mathcal{V}^i_{e_1(v)},\mathcal{V}^i_{e_2(v)})$ and the plane identified by the vectors $( \vec{ \ell}_{v,1}\vec{ \ell}_{v,2})$;
    \item[(2)]  $\vartheta_{v,2}$  as the angle between the plane identified by the vectors $ (\mathcal{V}^i_{e_3(v)},\mathcal{V}^i_{e_4(v)})$ and the plane identified by the vectors $( \vec{ \ell}_{v,1},  \vec{ \ell}_{v,2})$;
    \item[(3)]  $\vartheta_{v,3}$  as the angle between the plane identified by the vectors $ (\mathcal{V}^i_{e_5(v)},\mathcal{V}^i_{e_6(v)})$ and the plane identified by the vectors $( \vec{ \ell}_{v,3},  \vec{ \ell}_{v,4})$; and
    \item[(4)] $\vartheta_{v,4}$  as the angle between the plane identified by the vectors $( \vec{ \ell}_{v,1},  \vec{ \ell}_{v,2})$   and the plane identified by the vectors $( \vec{ \ell}_{v,3},  \vec{ \ell}_{v,4})$. 
\end{itemize}
{Moreover, we define $\breve{\ell}_v$ as the module of the Gauss constraint, i.e.,
\begin{equation}
\breve{\ell}_v:=\sqrt{\delta_{ij}G^i_vG^j_v},
\end{equation}} and introduce $\mathcal{V}_{v}\equiv\mathcal{V}_{v}^i\tau_i:=\frac{G^i_v\tau_i}{\breve{\ell}_v}$ .
{Then, another angle variable $\breve{\vartheta}_v$ can be defined as the angle between the plane identified by  the $su(2)$-valued vectors $({ \ell}^i_{v,3}\tau_i, \mathcal{V}_{v}^i\tau_i)$ and the plane identified by  $(n(\mathcal{V}_{v})\tau_2 n(\mathcal{V}_{v})^{-1}, \mathcal{V}_{v}^i\tau_i ) $,  see the illustration of this definition in Fig.\ref{fig:label2}.
It is easy to see that  $\breve{\vartheta}_v$  is conjugate to the module of Gauss constraint, which reads
  \begin{equation}
 \{\breve{\ell}_v,\breve{\vartheta}_{v'}\}=\delta_{v,v'}\frac{\kappa}{a^2}.
\end{equation}
}{The geometric interpretation of $\breve{\ell}_v \mathcal{V}^i_{v}$ and $\breve{\vartheta}_v$ are shown in Fig.\ref{fig:label2}.  Now, one can see that for fixed $(\eta_{e_1(v)},...,\eta_{e_6(v)})$, the unit vectors $(\mathcal{V}^i_{e_1(v)},...,\mathcal{V}^i_{e_6(v)})$ can be determined by $(\breve{\ell}_v,\breve{\vartheta}_{v'})$ and  $({ \ell}_{v,I}, \vartheta_{v',I}), \forall I\in\{1,2,3,4\}$ exactly. Now, the gauge reduction with respect to the Gauss constraint in $P_\gamma^+$ can be carried out as follows; As shown in Fig. \ref{fig:label222} , one should first impose the condition $\breve{\ell}_v=0$ to identify the two pairs $({ \ell}_{v,3}, \vartheta_{v,3})$ and $({ \ell}_{v,4}, \vartheta_{v,4})$ at each $v\in V(\gamma)$ , and then  reduce the gauge degrees of freedom carried by $\breve{\vartheta}_{v}$ and $\mathcal{V}_v$ to get the reduced twisted geometry space $H_\gamma^+$. As a summary, one can conclude that the new twisted geometric variables   $$(\eta_e,\varsigma_e),({ \ell}_{v,I}, \vartheta_{v,I})|_{I=1,2,3,4}, (\breve{\ell}_v,\breve{\vartheta}_{v}), \mathcal{V}_v$$provides a re-parametrization of the twisted geometry space $P_\gamma^+$. Specifically, the reduced twisted geometric variables   $(\eta_e,\varsigma_e),({ \ell}_{v,I}, \vartheta_{v,I})|_{I=1,2,3} $ parametrize the reduced twisted geometry space $H_\gamma^+$.

\begin{figure}[h]
 \centering
 \includegraphics[scale=0.25]{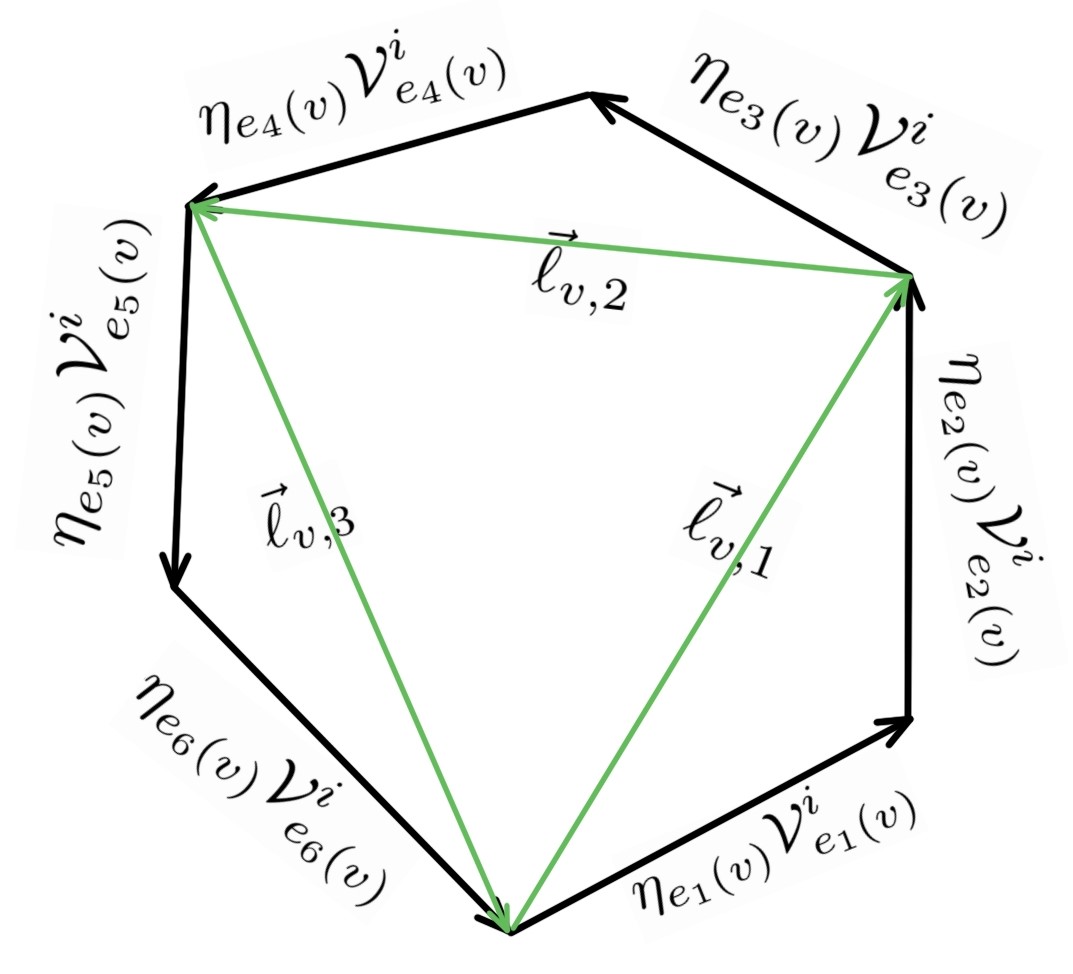}
\caption{The reduced twisted geometry associated to the vertex $v$.}
\label{fig:label222}
\end{figure}

It is necessary to clarify  the boundary $\bar{P}_\gamma^+$ and the interior $\dot{P}_\gamma^+$ of the twisted geometry space $P_\gamma^+=\dot{P}_\gamma^+\cup  \bar{P}_\gamma^+ $, where $\dot{P}_\gamma^+$ is composed by  the phase space points satisfying the following conditions,
 \begin{eqnarray}\label{boundary000}
 |\eta_{e_1(v)}-\eta_{e_2(v)}|<{ \ell}_{v,1}<\eta_{e_1(v)}+\eta_{e_2(v)},&&\ 
  \  |\eta_{e_3(v)}-\eta_{e_4(v)}|<{ \ell}_{v,2}<\eta_{e_3(v)}+\eta_{e_4(v)},\\\nonumber|\eta_{e_5(v)}-\eta_{e_6(v)}|<{ \ell}_{v,3}<\eta_{e_5(v)}+\eta_{e_6(v)},&&\  \  |{ \ell}_{v,1}-{ \ell}_{v,2}|<{ \ell}_{v,4}<{ \ell}_{v,1}+{ \ell}_{v,2},\\\nonumber
 |{ \ell}_{v,3}-\breve{ \ell}_{v}|<{ \ell}_{v,4}<{ \ell}_{v,3}+\breve{ \ell}_{v},&&\quad \forall v\in V(\gamma).
\end{eqnarray}
One can notice that each condition in  \eqref{boundary000} determines  a non-vanishing triangle in Fig.\ref{fig:label2}. Moreover, some of the triangles in  Fig.\ref{fig:label2} are vanished on the boundary $\bar{P}_\gamma^+$, and the vanishing triangles also reduces the degrees of freedom of the corresponding angles variables \cite{PhysRevLett.107.011301,Bianchi:2012wb}. Similarly, the reduced twisted geometry space can be also decomposed as boundary and interior as  $H_\gamma^+=\dot{H}_\gamma^+\cup \bar{H}_\gamma^+ $, where $\dot{H}_\gamma^+$ is composed by  the phase space points satisfying the following conditions,
 \begin{eqnarray}\label{boundary}
 |\eta_{e_1(v)}-\eta_{e_2(v)}|<{ \ell}_{v,1}<\eta_{e_1(v)}+\eta_{e_2(v)},&&\ 
  \  |\eta_{e_3(v)}-\eta_{e_4(v)}|<{ \ell}_{v,2}<\eta_{e_3(v)}+\eta_{e_4(v)},\\\nonumber|\eta_{e_5(v)}-\eta_{e_6(v)}|<{ \ell}_{v,3}<\eta_{e_5(v)}+\eta_{e_6(v)},&&\  \  |{ \ell}_{v,1}-{ \ell}_{v,2}|<{ \ell}_{v,3}<{ \ell}_{v,1}+{ \ell}_{v,2},\quad \forall v\in V(\gamma).
\end{eqnarray}
Now, the following theorem shows that the   symplectic structure on $\dot{P}_\gamma^+$ and $\dot{H}_\gamma^+$ can be expressed in terms of the new geometric  variables, and the proof can be found in Appendix \ref{app20}.
\begin{theorem}\label{theorem1}
     For the twisted geometry space $P_\gamma^+=(\times_{e\in E(\gamma)} T^\ast S_e^1)^+\times_{v\in V(\gamma)} \mathfrak{P}_v$ on a the cubic graph $\gamma$ on $\mathbb{T}^3$, where $(\eta_e,\varsigma_e)\in (T^\ast S_e^1)^+\cong \mathbb{R}^+_e\times S^1_e$ and $(({ \ell}_{v,I}, \vartheta_{v,I}), (\breve{\ell}_v,\breve{\vartheta}_{v}), \mathcal{V}_v^i)\in \mathfrak{P}_v, \quad I\in\{1,...,4\}$,
the symplectic 1-form on its interior $\dot{P}_\gamma^+$ of $P_\gamma^+$ can be re-expressed as
  \begin{equation}\label{sym3}
  \Theta_{\dot{P}_\gamma^+}=\frac{a^2}{\kappa}\left(\sum_{e\in\gamma}\eta_ed\varsigma_e+\sum_{v\in\gamma}\left(\breve{\ell}_v\text{Tr}(\mathcal{V}_vdn_vn_v^{-1}) +\sum_{I=1}^{4}\vartheta_{v,I}d{ \ell}_{v,I} +\breve{\vartheta}_{v}d\breve{\ell}_v\right)\right).
\end{equation}
By carrying out the symplectic reduction with respect to Gauss constraint, the reduced symplectic 1-form on the interior $\dot{H}_\gamma^+$ of the reduced phase space $H_\gamma^+$ can be given by
  \begin{equation}\label{sym4}
  \Theta_{\dot{H}_\gamma^+}=\frac{a^2}{\kappa}\left(\sum_{e\in\gamma}\eta_ed\varsigma_e+\sum_{v\in\gamma}\sum_{I=1}^{3}\vartheta_{v,I}d{ \ell}_{v,I} \right).
\end{equation}
\end{theorem}
}

\section{Twisted geometry coherent state and Symplectic structure}\label{sec:three}
\subsection{Basic property of twisted geometry coherent state}\label{sec:threeone} 
The twisted geometry coherent state can be regarded as a modification of the heat-kernel coherent state in $SU(2)$ LQG.Specifically, this is achieved by expanding the heat-kernel coherent state using twisted geometry parameters. The dominant contributions are then isolated, corresponding to the Perelomov coherent states $|j_e, j_e\rangle$ and $|j_e, -j_e\rangle$ of $SU(2)$, as outlined in \cite{Calcinari_2020}. Following this procedure, we get the twisted geometry coherent state:

\begin{eqnarray}\label{TGCS}
&&{\Psi}^t_{\gamma,\mathbb{H}}({h})=\prod_{e\in E(\gamma)}{\Psi}^t_{\mathbb{H}_e}(h_e)
\end{eqnarray}
 where $t:=\frac{\kappa\hbar}{a^2}$, $\mathbb{H}:=\{\mathbb{H}_e\}_{e\in\gamma}$ with ${\mathbb{H}}_e:=\{V_e,\tilde{V}_e,\xi_e,\eta_e\}$, and ${\Psi}^t_{\mathbb{H}_e}(h_e)$ reads
\begin{eqnarray}\label{twcs}
\nonumber{\Psi}^t_{\mathbb{H}_e}({h}_e)&=&\sum_{j_e} (\dim(\pi_{j_e})) \left(\exp(-\frac{t}{2}(\frac{\eta_e}{t}-d_{j_e})^2)e^{-\mathbf{i}\xi_ed_{j_e}} \langle j_e,j_e|n_e^{-1}h_e\tilde{n}_e|j_e,j_e\rangle\right.\\
&&+\left.\exp(-\frac{t}{2}(\frac{\eta_e}{t}+d_{j_e})^2)e^{\mathbf{i}\xi_ed_{j_e}}\langle j_e,-j_e|n_e^{-1}h_e\tilde{n}_e|j_e,-j_e\rangle\right).
\end{eqnarray}
Here, we define $\dim(\pi_{j_e})=2j_e+1$ and  $d_{j_e}\equiv (j_e+\frac{1}{2})$. Notably, the state ${n}_e|j_e,j_e\rangle$ involved in \eqref{twcs} is just the generalized Perelomov type coherent state of $SU(2)$, which can be also denoted by $|j_e,-\tilde{V}_e\rangle=\tilde{n}_e|j_e,j_e\rangle$ and $|j_e,{V}_e\rangle={n}_e|j_e,j_e\rangle$.
  The state  ${\Psi}^t_{\gamma,\mathbb{H}}({h})$ is gauge variant. It transforms as
 \begin{equation}
 {\Psi}^t_{\gamma,\mathbb{H}}({h})\mapsto {\Psi}^t_{\gamma,\mathbb{H}^{h'}}({h})= \prod_{e\in E(\gamma)}{\Psi}_{\mathbb{H}_e}(h'_{s(e)}h_eh'^{-1}_{t(e)})
 \end{equation}
 under the gauge transformation given by $h'=\{h'_v\}_{v\in V(\gamma)}$. The corresponding gauge invariant state can be give by taking group averaging over the $SU(2)$ gauge transformation.

Due to the tensor product structure shown in \eqref{TGCS},  we can focus on   ${\Psi}^t_{\mathbb{H}_e}$ to discuss the properties of the twisted geometry coherent state without loss of generality.
We first introduce the normalized version of ${\Psi}^t_{\mathbb{H}_e}$ as
\begin{eqnarray}
&&\tilde{\Psi}^t_{\mathbb{H}_e}({h}_e){=}\frac{{\Psi}^t_{\mathbb{H}_e}(h_e)} {||{\Psi}^t_{\mathbb{H}_e}||}
\end{eqnarray}
with
\begin{eqnarray}
&&||{\Psi}^t_{\mathbb{H}_e}||^2 \stackrel{\text{large}\  \eta_e}{=}\sqrt{\frac{\pi}{t}} \breve{\text{P}}(\frac{\eta_e}{t})\left(1+\mathcal{O}(e^{-\frac{1}{t}})+\mathcal{O}(\frac{t}{\eta_e})(1+\mathcal{O} (e^{-\frac{\eta_e^2}{2t}}))\right),
\end{eqnarray}
where the function $\breve{\text{P}}$ is defined as $\breve{\text{P}}(x):=2x+1$. Then, one can show that $\tilde{\Psi}^t_{\mathbb{H}_e}(h)$ provides a resolution of identity in $\mathcal{H}_e$ , which reads \cite{Calcinari_2020}
\begin{equation}\label{resoid1}
\mathbbm{1}_{\mathcal{H}_e}=
\int_{P_e^+}d\mathbb{H}_e|\tilde{\Psi}^t_{\mathbb{H}_e}\rangle\langle \tilde{\Psi}^t_{\mathbb{H}_e}|, 
\end{equation}
with
\begin{equation}
d\mathbb{H}_e:=||{\Psi}^t_{\mathbb{H}_e}||^2\frac{C(\eta_e)}{\sqrt{\pi t}}d\eta_e dV_ed\tilde{V}_e\frac{d\xi_{e}}{2\pi}=\frac{2\eta^2_e}{t^3}d\eta_e dV_ed\tilde{V}_e\frac{d\xi_{e}}{\pi}\left(1+\mathcal{O}(e^{-\frac{1}{t}})+\mathcal{O}(\frac{t}{\eta_e})\right),
\end{equation}
where $d\eta_e$ is the Lebesgue measure on $\mathbb{R}$, $\xi_e\in S_e^1:=[0, 2\pi)$, $C(\eta_e)=\frac{2\eta_e}{t}$ is determined by $\int_{\mathbb{R}}d\eta_e\frac{C(\eta_e)}{\sqrt{\pi t}}\exp(-t(\frac{\eta_e}{t}-d_{j_e})^2)=2j_e+1$, and $d{V}_{e}$, $d\tilde{V}_{e}$ are the normalized invariant measure on unit 2-sphere $S^2_e$ satisfying $\int_{S^2_e}d{V}_{e}=\int_{S^2_e}d\tilde{V}_{e}=1$.

It is also necessary to consider the overlap amplitude between two coherent states, which can be calculated as \cite{Long:2021lmd,Long:2022cex}
\begin{eqnarray}\label{prepsipsi}
&&\langle {\Psi}^t_{\mathbb{H}_e},{\Psi}^t_{\mathbb{H}'_e}\rangle\\\nonumber
&{=}&\sum_{N_e}\dim(\pi_{j_e})\exp(-\frac{t}{2}(\frac{\eta_e}{t}-d_{j_e})^2 -\frac{t}{2}(\frac{\eta'_e}{t}-d_{j_e})^2) \exp(- j_e\widetilde{\Omega}_e) e^{\mathbf{i}d_{j_e}(\xi_e-\xi'_e)} e^{\mathbf{i}j_e\tilde{\varphi}_e}\\\nonumber
&&+\frac{1}{\sqrt{t}}\mathcal{O}(e^{-\frac{\eta_e^2}{4t}})
\end{eqnarray}
for large $\eta_e$, where we defined $\widetilde{\Omega}_e:=\Omega(n_e,n'_e)+\Omega(\tilde{n}_e,\tilde{n}'_e)$, $\tilde{\varphi}_e:=\varphi(n_e,n'_e) +\varphi(\tilde{n}_e,\tilde{n}'_e)$ with $\Omega(n_e,n'_e)$, $\Omega(\tilde{n}_e,\tilde{n}'_e)$, $\varphi(n_e,n'_e)$ and $\varphi(\tilde{n}_e,\tilde{n}'_e)$ being defined by
  \begin{equation}
  \Omega(n_e,n'_e):=-\frac{\ln|\langle j_e, V'_e|j_e, V_e\rangle|}{j_e}=-\ln|\langle 1, V'_e|1, V_e\rangle|\geq0,
  \end{equation}
  \begin{equation}
  \Omega(\tilde{n}_e,\tilde{n}'_e):=-\frac{\ln|\langle j_e, -\tilde{V}_e|j_e,-\tilde{ V}'_e\rangle|}{j_e} =-\ln|\langle 1, -\tilde{V}_e|1,-\tilde{ V}'_e\rangle|\geq0,
  \end{equation}
   \begin{equation}
   e^{\mathbf{i}j_e\varphi(n_e,n'_e)}:=\frac{\langle j_e, V'_e|j_e, V_e\rangle}{|\langle j_e, V'_e|j_e, V_e\rangle|},
   \end{equation}
   and
   \begin{equation}
   e^{\mathbf{i}j_e\varphi(\tilde{n}_e,\tilde{n}'_e)}:=\frac{\langle j_e, -\tilde{V}_e|j_e, -\tilde{V}'_e\rangle}{|\langle j_e, -\tilde{V}_e|j_e, -\tilde{V}'_e\rangle|}
   \end{equation}
respectively. Then, by using Poisson summation formulation,  one can give the result of the overlap amplitude as \cite{Long:2021lmd,Long:2022cex}
 \begin{equation}\label{ZHH}
\langle \tilde{\Psi}^t_{\mathbb{H}_e},\tilde{\Psi}^t_{\mathbb{H}'_e}\rangle \simeq
\exp(\mathfrak{I}(\mathbb{H}_e, \mathbb{H}'_e)/t),
\end{equation}
with
\begin{eqnarray}
\mathfrak{I}(\mathbb{H}_e, \mathbb{H}'_e)&:=& -\frac{1}{2}(\eta'_e-{\eta_e})^2 +(\frac{\eta'_e}{2}-\frac{\eta_e}{2}-\frac{\widetilde{\Omega}_e}{2})^2-\frac{(\xi_e-\xi'_e+\tilde{\varphi}_e)^2}{4}\\\nonumber
&&+\mathbf{i}(\frac{\eta_e}{2}+\frac{\eta'_e}{2}-\frac{\widetilde{\Omega}_e}{2}) (\xi_e-\xi'_e+\tilde{\varphi}_e)-\eta_e\widetilde{\Omega}_e.
\end{eqnarray}

\subsection{Symplectic structure and canonical transformation for twisted geometry}

Let us use complex twisted geometry parameters  to re-express the previous analysis for twisted geometry coherent state. Given a set of real twisted geometry parameters $\eta_e,\xi_e, V_e$ and $\tilde V_e$, the complex twisted geometry parameters are defined by
\begin{equation}\label{fromreal2complex}
u_e:=\frac{1}{\sqrt{2}}({\eta_e}-\mathbf{i}\xi_e),\quad v_e:=\frac{1}{\sqrt{2}}({\eta_e}+\mathbf{i}\xi_e), 
\end{equation}
\begin{equation}\label{fromreal2complex22}
z_e:=\exp(\mathbf{i}\phi_e)\cot\frac{\theta_e}{2},\quad w_e:=\exp(-\mathbf{i}\phi_e)\cot\frac{\theta_e}{2},
\end{equation}
\begin{equation}\label{fromreal2complex33}
\tilde{z}_e:=\exp(-\mathbf{i}\tilde{\phi}_e)\cot\frac{\tilde{\theta}_e}{2},\quad \tilde{w}_e:=\exp(\mathbf{i}\tilde{\phi}_e)\cot\frac{\tilde{\theta}_e}{2}
\end{equation}
where $(\theta_e,\phi_e)$ and   $(\tilde{\theta}_e,\tilde{\phi}_e)$ is the spherical coordinates of $V_e$ and $-\tilde{V}_e$ respectively.  For convenience, let us  use $\mathbb C(e)$ to denote the space coordinated by $(u_e,z_e,\tilde z_e)$, and use $\mathbb C^*(e)$ to denote the one coordinated by $(v_e,w_e,\tilde{w}_e)$. It is noted  that, by definition, $\mathbb C^*(e)$ is the complex conjugate of $\mathbb C(e)$. Due to the one-to-one correspondence between $\mathbb H_e$ and $\mathbb C(e)$, the coherent state $\tilde \Psi_{\mathbb H_e}^t$, as well as ${\Psi}_{\mathbb H_e}^t$ defined in Sec. \ref{sec:threeone} can be denoted by $\tilde \Psi_{\mathbb C(e)}^t$, as well as  ${\Psi}_{\mathbb C(e)}^t$ respectively. 
Under the transformation given by Eqs. \eqref{fromreal2complex}, \eqref{fromreal2complex22} and  \eqref{fromreal2complex33}, the measures transform as
\begin{equation}
dV_e\to\frac{1}{2\mathbf{i}\pi}\frac{dw_e d z_e}{(1+w_e z_e)^2},\quad d\tilde{V}_e\to\frac{1}{2\pi\mathbf{i}}\frac{d\tilde{w}_e d\tilde{z}_e}{(1+\tilde{w}_e \tilde{z}_e)^2},\quad d\eta_ed\xi_{e}\to \frac{du_edv_e}{\mathbf{i}}.
\end{equation}
Correspondingly, the resolution of identity can be reformulated as 
\begin{equation}\label{resoid2}
\mathbbm{1}_{\mathcal{H}_e}=
\int_{}d\nu(\mathbb{C}^\ast(e),\mathbb{C}(e))|\tilde{\Psi}^t_{\mathbb{C}(e)}\rangle\langle \tilde{\Psi}^t_{\mathbb{C}(e)}|, %\quad \sigma^0_e\sim \sqrt{N_e},
\end{equation}
where  the measure $d\nu(\mathbb{C}^\ast(e),\mathbb{C}(e))$ is defined by
\begin{eqnarray}\label{mea1}
d\nu(\mathbb{C}^\ast(e),\mathbb{C}(e))&:=&C(\eta_e)||{\Psi}^t_{\mathbb{C}_e}||^2\frac{d\eta_e}{\sqrt{\pi t}} dV_ed\tilde{V}_e\frac{d\xi_{e}}{2\pi}\\\nonumber
&=&\frac{du_edv_e}{2\pi t\mathbf{i}}\frac{\eta_e d\tilde{w}_ed\tilde{z}_e}{\pi t\mathbf{i}(1+\tilde{w}_e \tilde{z}_e)^2}\frac{\eta_e dw_e d z_e}{\pi t\mathbf{i}(1+w_e z_e)^2}(1+\mathcal{O}(\frac{t}{\eta})+\mathcal{O}(e^{-\frac{1}{t}})).
\end{eqnarray}

The inner product \eqref{ZHH} between two twisted geometry  coherent state can also be expressed via the complex coordinate of twisted geometry parameter space. The result is
 \begin{equation}\label{ZHH2}
\langle \tilde{\Psi}^t_{\mathbb{C}(e)},\tilde{\Psi}^t_{\mathbb{C}'(e)}\rangle \simeq
\exp(\mathfrak{I}(\mathbb{C}(e),\mathbb{C}^\ast(e); \mathbb{C}'(e), \mathbb{C}'^\ast(e))/t)
\end{equation}
where $\mathfrak{I}(\mathbb{C}(e),\mathbb{C}^\ast(e); \mathbb{C}'(e), \mathbb{C}'^\ast(e))$  is given by
\begin{eqnarray}\label{ZHH3}
&&\mathfrak{I}(\mathbb{C}(e),\mathbb{C}^\ast(e); \mathbb{C}'(e), \mathbb{C}'^\ast(e))\nonumber\\
%&:=&-\frac{1}{2}(\eta'(e)-\eta(e))^2 +\frac{1}{4}(\eta'(e)-\eta(e)-\widetilde{\Omega}(e))^2-\frac{(\xi(e)-\xi'(e)+\tilde{\varphi}(e))^2}{4}\\\nonumber
%&&+\mathbf{i}(\frac{\eta(e)}{2}+\frac{\eta'(e)}{2}-\frac{\widetilde{\Omega}(e)}{2}) (\xi(e)-\xi'(e)+\tilde{\varphi}(e))-\eta(e)\widetilde{\Omega}(e)\\\nonumber
&=&-\frac{v(e)u(e)}{2}+{v(e)u'(e)}-\frac{v'(e)u'(e)}{2}+\frac{\mathbf{i}\eta(e)\xi(e)-\mathbf{i}\eta'(e)\xi'(e)}{4}\\\nonumber
&&-\frac{1}{\sqrt{2}}(v(e)+u'(e))(\widetilde{\Omega}(e)-\mathbf{i}\tilde{\varphi}(e))+\frac{1}{4}(\widetilde{\Omega}(e)-\mathbf{i}\tilde{\varphi}(e))^2\label{ICC},
\end{eqnarray}
with
\begin{eqnarray}
\widetilde{\Omega}(e)-\mathbf{i}\tilde{\varphi}(e)&\equiv&\Omega(n_e,n'_e)-\mathbf{i}\varphi(n_e,n'_e)+\Omega(\tilde{n}_e,\tilde{n}'_e)-\mathbf{i}\varphi(\tilde{n}_e,\tilde{n}'_e)\\\nonumber
&=&-\ln(\frac{(1+w_ez'_e)^2}{(1+w_ez_e)(1+w'_ez'_e)})-\ln(\frac{(1+\tilde{w}_e\tilde{z}'_e)^2}{(1+\tilde{w}_e\tilde{z}_e)(1+\tilde{w}'_e\tilde{z}'_e)}).
\end{eqnarray} 
{Notably, the term $\widetilde{\Omega}(e)-\mathbf{i}\tilde{\varphi}(e)$ is involved in \eqref{ICC} due to
 \begin{eqnarray}\label{VVwz1}
\langle 1, V'_e|1, V_e\rangle=\exp{(-\Omega(n_e,n'_e)+\mathbf{i}\varphi(n_e,n'_e))}=\frac{(1+w_ez'_e)^2}{(1+w_ez_e)(1+w'_ez'_e)}
\end{eqnarray}
and 
\begin{eqnarray}\label{VVwz2}
\langle 1, -\tilde{V}_e|1,-\tilde{ V}'_e\rangle= \exp{(-\Omega(\tilde{n}_e,\tilde{n}'_e)+\mathbf{i}\varphi(\tilde{n}_e,\tilde{n}'_e))}=\frac{(1+\tilde{w}_e\tilde{z}'_e)^2}{(1+\tilde{w}_e\tilde{z}_e)(1+\tilde{w}'_e\tilde{z}'_e)}
\end{eqnarray}
for the Perelomov type coherent state of  $SU(2)$.}
 {As what we did for $\mathbb H_e$ in \eqref{TGCS}, let us introduce the notion $\mathbb{C}=\{\mathbb C(e)\}_{e\in\gamma}$. Then the coherent state $\tilde\Psi_{\gamma,\mathbb H}^t$ associated with the entire graph can be relabelled by }{$\tilde\Psi_{\gamma,\mathbb C}^t$. Since the cubic graph $\gamma$ on $\mathbb{T}^3$ will be focused on in what follows, we thus drop the index $\gamma$ and simply denote the coherent state as $\tilde\Psi_{\mathbb C}^t$.  }
{Then, given two coherent states $\tilde{\Psi}_{\mathbb C}^t$ and $\tilde{\Psi}_{\mathbb C'}^t$ corresponding to two sets of parameters $\mathbb{C}$ and $\mathbb{C}'$ respectively, there inner product is}
\begin{equation}\label{ZHH3C}
\langle \tilde{\Psi}^t_{\mathbb{C}},\tilde{\Psi}^t_{\mathbb{C}'}\rangle \simeq
\exp(\mathfrak{I}(\mathbb{C},\mathbb{C}^\ast; \mathbb{C}', \mathbb{C}'^\ast)/t),
\end{equation}
{with $$\mathfrak{I}(\mathbb{C},\mathbb{C}^\ast; \mathbb{C}', \mathbb{C}'^\ast):=\sum_{e\in\gamma}\mathfrak{I}(\mathbb{C}(e),\mathbb{C}^\ast(e); \mathbb{C}'(e), \mathbb{C}'^\ast(e)).
$$}

{To align with the notation used in the next section, we will re-denote $\mathbb C$ and  $\mathbb C^*$ as $\mathbb{C}_{\imath+1}$ and  $\mathbb{C}^*_{\imath+1}$ respectively, and redenote $\mathbb{C}'$ and $\mathbb{C}'^\ast$ as $\mathbb{C}_{\imath}$ and  $\mathbb{C}^*_{\imath}$. Then, we introduce the Kahler potential  $\mathfrak{K}(\mathbb{C}_{\imath+1},\mathbb{C}_{\imath+1}^\ast; \mathbb{C}_{\imath}, \mathbb{C}_{\imath}^\ast)/t$ by defining}
\begin{eqnarray}\label{KIrelation}
&&\mathfrak{K}(\mathbb{C}_{\imath+1},\mathbb{C}_{\imath+1}^\ast; \mathbb{C}_{\imath}, \mathbb{C}_{\imath}^\ast)\\\nonumber
&:=&\mathfrak{I}(\mathbb{C}_{\imath+1},\mathbb{C}_{\imath+1}^\ast; \mathbb{C}_{\imath}, \mathbb{C}_{\imath}^\ast)-\sum_{e}\frac{\mathbf{i}\eta_e^{\imath+1}\xi_e^{\imath+1}-\mathbf{i}\eta_e^{\imath}\xi_e^{\imath}}{4}\\\nonumber
&&-\frac{1}{4}\sum_{e}\left(\Omega(n^{\imath+1}_e,n^{\imath}_e)-\mathbf{i}\varphi(n^{\imath+1}_e,n^{\imath}_e)+\Omega(\tilde{n}^{\imath+1}_e,\tilde{n}^{\imath}_e)-\mathbf{i}\varphi(\tilde{n}^{\imath+1}_e,\tilde{n}^{\imath}_e)\right)^2\\\nonumber
&=&\sum_{e}-\frac{v^{\imath+1}_eu^{\imath+1}_e}{2}+v^{\imath+1}_eu^\imath_e-\frac{v^\imath_eu^\imath_e}{2}\\\nonumber
&&+{\sqrt{2}}(v^{\imath+1}_e+u^\imath_e)\left(\ln(1+w^{\imath+1}_ez^{\imath}_e)-\frac{1}{2}\ln(1+w^{\imath+1}_ez^{\imath+1}_e)-\frac{1}{2}\ln(1+w^\imath_ez^\imath_e)\right)
\\\nonumber
&&+{\sqrt{2}}(v^{\imath+1}_e+u^\imath_e)\left(\ln(1+\tilde{w}^{\imath+1}_e\tilde{z}^{\imath}_e)-\frac{1}{2}\ln(1+\tilde{w}^{\imath+1}_e\tilde{z}^{\imath+1}_e)-\frac{1}{2}\ln(1+\tilde{w}^{\imath}_e\tilde{z}^{\imath}_e)\right),
\end{eqnarray}
{which is equal to $\mathfrak{I}(\mathbb{C}_{\imath+1},\mathbb{C}_{\imath+1}^\ast; \mathbb{C}_{\imath}, \mathbb{C}_{\imath}^\ast)/t$ up to some boundary terms and higher order terms.}
The Kahler potential  $\mathfrak{K}(\mathbb{C}_{\imath+1},\mathbb{C}_{\imath+1}^\ast; \mathbb{C}_{\imath}, \mathbb{C}_{\imath}^\ast)/t$ can be related to the symplectic one-form on the phase space $\dot{P}_\gamma^+$ as follows.  
{Consider a curve $\tau\mapsto \mathbb C(\tau)$ in the phase space $\dot{P}_\gamma^+$. Choose  two points $\mathbb C_{\imath+1}$ and $\mathbb C_{\imath}$ in the curve with $\mathbb C_{\imath}\equiv \mathbb C(\tau)$ and $\mathbb C_{\imath+1}\equiv \mathbb C(\tau+\Delta\tau)$. Then, one can check that
}
\begin{eqnarray}\label{oneform1}
&&\lim_{\Delta\tau\to0}\sum_{e} \Bigg(\frac{\partial \mathfrak{K}(\mathbb{C}_{\imath+1},\mathbb{C}_{\imath+1}^\ast; \mathbb{C}_\imath, \mathbb{C}_\imath^\ast)}{\partial v_e^{\imath+1}}\frac{(v_e^{\imath+1}-v_e^\imath)}{\Delta \tau}-\frac{\partial \mathfrak{K}(\mathbb{C}_{\imath+1},\mathbb{C}_{\imath+1}^\ast; \mathbb{C}_\imath, \mathbb{C}_\imath^\ast)}{\partial u_e^\imath}\frac{(u_e^{\imath+1}-u_e^\imath)}{\Delta \tau}\\\nonumber
&&+\frac{\partial \mathfrak{K}(\mathbb{C}_{\imath+1},\mathbb{C}_{\imath+1}^\ast; \mathbb{C}_\imath, \mathbb{C}_\imath^\ast)}{\partial w_e^{\imath+1}}\frac{(w_e^{\imath+1}-w_e^\imath)}{\Delta \tau}-\frac{\partial \mathfrak{K}(\mathbb{C}_{\imath+1},\mathbb{C}_{\imath+1}^\ast; \mathbb{C}_\imath, \mathbb{C}_\imath^\ast)}{\partial z_e^\imath}\frac{(z_e^{\imath+1}-z_e^\imath)}{\Delta \tau}\\\nonumber
&&+\frac{\partial \mathfrak{K}(\mathbb{C}_{\imath+1},\mathbb{C}_{\imath+1}^\ast; \mathbb{C}_\imath, \mathbb{C}_\imath^\ast)}{\partial \tilde{w}_e^{\imath+1}}\frac{(\tilde{w}_e^{\imath+1}-\tilde{w}_e^\imath)}{\Delta \tau}-\frac{\partial \mathfrak{K}(\mathbb{C}_{\imath+1},\mathbb{C}_{\imath+1}^\ast; \mathbb{C}_\imath, \mathbb{C}_\imath^\ast)}{\partial \tilde{z}_e^\imath}\frac{(\tilde{z}_e^{\imath+1}-\tilde{z}_e^\imath)}{\Delta \tau}\Bigg)\\\nonumber
&=&\sum_{e}\Bigg(\frac{1}{2}(u_e\frac{d{v}_e}{d\tau}-v_e\frac{d{u}_e}{d\tau})+\frac{1}{\sqrt{2}}(u_e+v_e)(\frac{z_e \frac{d{w}_e}{d\tau}-w_e \frac{d{z}_e}{d\tau}}{1+z_e w_e}+\frac{\tilde{z}_e \frac{d{\tilde{w}}_e}{d\tau}-\tilde{w}_e \frac{d{\tilde{z}}_e}{d\tau}}{1+\tilde{w}_e \tilde{z}_e})\Bigg)\\\nonumber
&=&\frac{t}{\hbar}\frac{\partial}{\partial \tau}\circ \Theta_{\dot{P}_\gamma^+},
\end{eqnarray}
{where $\Theta_{\dot{P}_\gamma^+}$ is given by Eq.  \eqref{sym2} or \eqref{sym3}, and $\circ$ denotes the contraction of the vector field $\frac{\partial}{\partial \tau}$ and the one-form field $ \Theta_{\dot{P}_\gamma^+}$ on the phase space.}   Let us introduce the auxiliary relation
\begin{eqnarray}\label{KexpanC}
0&=&\mathfrak{K}(\mathbb{C}_{\imath+1},\mathbb{C}_{\imath+1}^\ast; \mathbb{C}_{\imath}, \mathbb{C}_{\imath}^\ast)|_{\mathbb{C}_{\imath}=\mathbb{C}_{\imath+1},\mathbb{C}_{\imath+1}^\ast=\mathbb{C}_{\imath}^\ast}\\\nonumber
&=&\mathfrak{K}(\mathbb{C}_{\imath+1},\mathbb{C}_{\imath+1}^\ast; \mathbb{C}_\imath, \mathbb{C}_\imath^\ast)\\\nonumber
&&+\sum_{e}\Bigg(\frac{\partial {\mathfrak{K}}(\mathbb{C}_{\imath+1},\mathbb{C}_{\imath+1}^\ast; \mathbb{C}_\imath, \mathbb{C}_\imath^\ast)}{\partial u_e^\imath}{(u_e^{\imath+1}-u_e^\imath)}-\frac{\partial {\mathfrak{K}}(\mathbb{C}_{\imath+1},\mathbb{C}_{\imath+1}^\ast; \mathbb{C}_\imath, \mathbb{C}_\imath^\ast)}{\partial v_e^{\imath+1}}{(v_e^{\imath+1}-v_e^\imath)}\\\nonumber
&&+\frac{\partial {\mathfrak{K}}(\mathbb{C}_{\imath+1},\mathbb{C}_{\imath+1}^\ast; \mathbb{C}_\imath, \mathbb{C}_\imath^\ast)}{\partial \tilde{z}_e^\imath}{(\tilde{z}_e^{\imath+1}-\tilde{z}_e^\imath)}-\frac{\partial {\mathfrak{K}}(\mathbb{C}_{\imath+1},\mathbb{C}_{\imath+1}^\ast; \mathbb{C}_\imath, \mathbb{C}_\imath^\ast)}{\partial \tilde{w}_e^{\imath+1}}{(\tilde{w}_e^{\imath+1}-\tilde{w}_e^\imath)}\\\nonumber
&&+\frac{\partial {\mathfrak{K}}(\mathbb{C}_{\imath+1},\mathbb{C}_{\imath+1}^\ast; \mathbb{C}_\imath, \mathbb{C}_\imath^\ast)}{\partial {z}_e^\imath}{({z}_e^{\imath+1}-{z}_e^\imath)}-\frac{\partial {\mathfrak{K}}(\mathbb{C}_{\imath+1},\mathbb{C}_{\imath+1}^\ast; \mathbb{C}_\imath, \mathbb{C}_\imath^\ast)}{\partial {w}_e^{\imath+1}}{({w}_e^{\imath+1}-{w}_e^\imath)}\Bigg)\\\nonumber
&&+\mathcal{O}((\Delta\tau)^2).
\end{eqnarray}
Combining this equation with Eq. \eqref{oneform1}, we can get
\begin{eqnarray}\label{oneformCcontinuum}
&&\lim_{\Delta\tau\to0}\frac{{\mathfrak{K}}(\mathbb{C}_{\imath+1},\mathbb{C}_{\imath+1}^\ast; \mathbb{C}_\imath, \mathbb{C}_\imath^\ast)}{\Delta \tau}\\\nonumber
&=&\sum_{e}\frac{1}{2}(u_e\frac{d{v}_e}{d\tau}-v_e\frac{d{u}_e}{d\tau})+\frac{1}{\sqrt{2}}(u_e+v_e)(\frac{z_e \frac{d{w}_e}{d\tau}-w_e \frac{d{z}_e}{d\tau}}{1+z_e w_e}+\frac{\tilde{z}_e \frac{d{\tilde{w}}_e}{d\tau}-\tilde{w}_e \frac{d{\tilde{z}}_e}{d\tau}}{1+\tilde{w}_e \tilde{z}_e})\\\nonumber
&=&\frac{t}{\hbar}\frac{\partial}{\partial \tau}\circ \Theta_{\dot{P}_\gamma^+}.
\end{eqnarray}

Since the new twisted geometry variables { give} a simpler {Poisson algebra structure} on the reduced phase space,  let us introduce the   new complex coordinate  $\mathbb{B}(e):=(Z_e,Z_{v,I},\check{Z}_v,\breve{Z}_v)$ and $\mathbb{B}^\ast(e):=(W_e,W_{v,I},\check{W}_v,\breve{W}_v)$ based on the new twisted geometry variables of the phase space  $P_\gamma^+$.{They are defined by}
\begin{equation}
Z_e:=\frac{1}{\sqrt{2}}(\eta_e-\mathbf{i}\varsigma_e),\quad W_e:=\frac{1}{\sqrt{2}}(\eta_e+\mathbf{i}\varsigma_e), 
\end{equation}
\begin{equation}
Z_{v,I}:=\frac{1}{\sqrt{2}}(\ell_{v,I}-\mathbf{i}\vartheta_{v,I}),\quad W_{v,I}:=\frac{1}{\sqrt{2}}(\ell_{v,I}+\mathbf{i}\vartheta_{v,I}), 
\end{equation}
\begin{equation}
\check{Z}_v:=\exp(\mathbf{i}\check{\phi}_v)\cot\frac{\check{\theta}_v}{2},\quad \check{W}_v:=\exp(-\mathbf{i}\check{\phi}_v)\cot\frac{\check{\theta}_v}{2},
\end{equation}
\begin{equation}
\breve{Z}_v:=\frac{1}{\sqrt{2}}(\breve{\ell}_v-\mathbf{i}\breve{\vartheta}_v),\quad \breve{W}_v:=\frac{1}{\sqrt{2}}(\breve{\ell}_v+\mathbf{i}\breve{\vartheta}_v),
\end{equation}
where $(\check{\theta}_v,\check{\phi}_v)$ is the spherical coordinates of the unit vector $\mathcal{V}^i_v$ .
Then, it is directly to express Eq.\eqref{oneformCcontinuum} as
\begin{eqnarray}\label{oneformBcontinuum}
&&\lim_{\Delta\tau\to0}\frac{{\mathfrak{K}}(\mathbb{B}_{\imath+1},\mathbb{B}_{\imath+1}^\ast; \mathbb{B}_\imath, \mathbb{B}_\imath^\ast)}{\Delta \tau}\\\nonumber
&=& \sum_{e}\frac{1}{2}(Z_e\frac{d{W}_e}{d\tau}-W_e \frac{d{Z}_e}{d\tau})+\sum_{v}\frac{1}{\sqrt{2}}(\breve{Z}_v+\breve{W}_v)(\frac{\check{Z}_v \frac{d\check{W}_v}{d\tau}-\check{W}_v\frac{ d\check{Z}_v}{d\tau}}{1+\check{Z}_v \check{W}_v})\\\nonumber
&&+\sum_{v}\sum_{I=1}^{4}\frac{1}{2}({{Z}_{v,I} \frac{d{W}_{v,I}}{d\tau}-\frac{d{Z}_{v,I}}{d\tau} {W}_{v,I}})+\sum_{v}\frac{1}{2}(\breve{Z}_v\frac{d\breve{W}_v}{d\tau}-\breve{W}_v\frac{ d\breve{Z}_v}{d\tau})\\\nonumber
&=&\frac{t}{\hbar}\frac{\partial}{\partial \tau}\circ \Theta_{\dot{P}_\gamma^+},
\end{eqnarray}
where we used the expression \eqref{sym3} for $\Theta_{\dot{P}_\gamma^+}$, and $\mathfrak{K}(\mathbb{B}_{\imath+1},\mathbb{B}_{\imath+1}^\ast; \mathbb{B}_\imath, \mathbb{B}_\imath^\ast)$ is given by re-expressing $\mathfrak{K}(\mathbb{C}_{\imath+1},\mathbb{C}_{\imath+1}^\ast; \mathbb{C}_{\imath}, \mathbb{C}_{\imath}^\ast)$  in terms of  the   new complex coordinate  $\mathbb{B}(e)$. 
Additionally,  since  the measure \eqref{mea1} is also adapted to  the symplectic one-form $\Theta_{\dot{P}_\gamma^+}$ on   $\dot{P}_\gamma^+$ , it can be reformulated by the new complex coordinate  $\mathbb{B}(e)$  as
\begin{eqnarray}\label{mea3B}
&&\prod_{e}d\nu(\mathbb{B}^\ast(e),\mathbb{B}(e))\\\nonumber
&=&\prod_{e}\frac{dZ_edW_e}{2t\pi\mathbf{i}}\prod_{v}\frac{d\breve{Z}_vd\breve{W}_v}{2t\pi\mathbf{i}}\frac{(\breve{Z}_v+\breve{W}_v)}{\sqrt{2}t}\frac{d^2\check{Z}_v}{\pi(1+\check{Z}_v \check{W}_v)^2}\prod_{I=1}^{4}\frac{dZ_{v,I}dW_{v,I}}{2t\pi\mathbf{i}}(1+\mathcal{O}(\frac{t}{\eta})+\mathcal{O}(e^{-\frac{1}{t}})),
\end{eqnarray}
where one should notice that $|E(\gamma)|=\frac{1}{3}|V(\gamma)|$ for the cubic graph $\gamma$ on the spatial manifold with topology $\mathbb{T}^3$.

\section{Path-integral and semiclassical propagator for twisted geometry coherent-state}\label{sec:four}

The dynamics of LQG can be defined by the physical Hamiltonian which is introduced in gravity coupled to some matter fields for deparametrization.
 In the deparametrization models with certain dust fields  where the dust reference frame is applied to provide the physical spatial coordinates and  time $\tau$, the scalar and diffeomorphism constraints are solved classically so that the theory can be described in terms of Dirac observables.
 The evolution with respect to the dust time is generated by the physical Hamiltonian \cite{Han_2020,Brown:1994py,PhysRevD.43.419,Domagala:2010bm}. 
 In the  deparametrization model  with the Gaussian dust \cite{PhysRevD.43.419}, the (non-graph-changing) physical Hamiltonian operator $\hat{\mathbf{H}}$ determining the quantum dynamics in $\mathcal{H}_\gamma$ can be given {as \cite{PhysRevD.43.419,Han_2020}
\begin{equation}\label{Hami}
\hat{\mathbf{H}}=\frac{1}{2}(\hat{\mathcal{C}}{[1]}+\hat{\mathcal{C}}{[1]}^\dagger),
\end{equation}
where $\hat{\mathcal{C}}{[1]}$ is the gravitational contribution to the scalar constraint operator with lapse function $\mathcal{N}=1$ \cite{Zhang:2021qul,Han_2020}. Based on this Hamiltonian operator, let us start to construct the coherent state path integral formulation on twisted geometry.}

The construction of the coherent state path integral formulation starts with  the transition amplitude associated with the unitary evolution generated by the physical Hamiltonian operator. Specifically, let us consider the transition amplitude
{
\begin{equation}
\int dg_1dg_2 \langle\tilde{\Psi}_{\mathbb{B}_f}^t|g_1U(T) g_2^{-1}|\tilde{\Psi}_{\mathbb{B}_i}^t\rangle=\int dg \langle\tilde{\Psi}_{\mathbb{B}_f}^t|U(T) g^{-1}|\tilde{\Psi}_{\mathbb{B}_i}^t\rangle
%\frac{q}{\sqrt{\int dg' \langle\tilde{\Psi}_{\mathbb{B}_f}^t|g'^{-1}|\tilde{\Psi}_{\mathbb{B}_f}^t\rangle\int dg'' \langle\tilde{\Psi}_{\mathbb{B}_i}^t|g''^{-1}|\tilde{\Psi}_{\mathbb{B}_i}^t\rangle}}
\end{equation}
}
between two gauge invariant coherent states  $\int dg g^{-1} |\tilde{\Psi}_{\mathbb{B}_i}^t\rangle$ and ${\int dg g^{-1} |\tilde{\Psi}_{\mathbb{B}_f}^t\rangle}$,
%  $\frac{\int dg g^{-1} |\tilde{\Psi}_{\mathbb{B}_i}^t\rangle}{\sqrt{\int dg' \langle\tilde{\Psi}_{\mathbb{B}_i}^t|g'^{-1}|\tilde{\Psi}_{\mathbb{B}_i}^t\rangle}}$ and $\frac{\int dg g^{-1} |\tilde{\Psi}_{\mathbb{B}_f}^t\rangle}{\sqrt{\int dg' \langle\tilde{\Psi}_{\mathbb{B}_f}^t|g'^{-1}|\tilde{\Psi}_{\mathbb{B}_f}^t\rangle}}$,
where $g=\{g_v|v\in\gamma\}$, and  the  initial state $\tilde{\Psi}_{\mathbb{B}_i}^t$ labelled by corner mark $i$ and final state $\tilde{\Psi}_{\mathbb{B}_f}^t$ labelled by corner mark $f$  are given by the twisted geometry coherent states. Let us denote $|\tilde{\Psi}_{\mathbb{B}^g_i}^t\rangle=g^{-1}|\tilde{\Psi}_{\mathbb{B}_i}^t\rangle$ with $\mathbb{B}^g_i$ given by transporting the phase space point  $\mathbb{B}_i$ by $g=\{g_v|v\in\gamma\}$. Then,
by inserting the resolution of identity \eqref{resoid2} with the measure \eqref{mea3B} to divide the  evolution into $N$ segments and using the inner product \eqref{ZHH3C}, the path-integral formulation of the propagator for gauge invariant  twisted geometry coherent-state  is given by
\begin{eqnarray}\label{AHH2}
% \nonumber to remove numbering (before each equation)
{K}(\mathbb{B}_f,\mathbb{B}_i,T)&=&\int dg\langle\tilde{\Psi}_{\mathbb{B}_f}^t|U(T) |\tilde{\Psi}_{\mathbb{B}^g_i}^t\rangle = \int dg\langle\tilde{\Psi}_{\mathbb{B}_f}^t|\left(e^{-\frac{\mathbf{i}}{\hbar}\Delta\tau\hat{\mathbf{H}} }\right)^{N}|\tilde{\Psi}_{\mathbb{B}^g_i}^t\rangle \\\nonumber
   &=& \int dg\prod_{\imath=1}^{N}d\nu[\mathbb{B}^\ast_\imath,\mathbb{B}_\imath]\langle\tilde{\Psi}_{\mathbb{B}_f}^t|  e^{-\frac{\mathbf{i}}{\hbar}\Delta\tau\hat{\mathbf{H}} }| \tilde{\Psi}_{\mathbb{B}_{N}}^t\rangle  \langle\tilde{\Psi}_{\mathbb{B}_{N}}^t| e^{-\frac{\mathbf{i}}{\hbar}\Delta\tau\hat{\mathbf{H}} }| \tilde{\Psi}_{\mathbb{B}_{N-1}}^t\rangle\\\nonumber
    &&...   \langle\tilde{\Psi}_{\mathbb{B}_{2}}^t| e^{-\frac{\mathbf{i}}{\hbar}\Delta\tau\hat{\mathbf{H}} }|\tilde{\Psi}_{\mathbb{B}_{1}}^t\rangle\langle\tilde{\Psi}_{\mathbb{B}_{1}}^t| e^{-\frac{\mathbf{i}}{\hbar}\Delta\tau\hat{\mathbf{H}} }|\tilde{\Psi}_{\mathbb{B}^g_i}^t\rangle\\\nonumber
 &=&\int dg\prod_{\imath=1}^{N}\prod_{e}d\nu(\mathbb{B}^\ast_\imath(e),\mathbb{B}_\imath(e))\exp(\mathbf{i}{S}[\mathbb{B}^\ast, \mathbb{B}]/t),
\end{eqnarray}
where we used the notations $\Delta \tau=T/N$, $[\mathbb{B}]:=\{\mathbb{B}_\imath|\imath\in\{0,1,...,N+1\}\}$, $\mathbb{B}_\imath=\{\mathbb{B}_\imath(e)|e\in\gamma\}$, and $\mathbb{B}_{N+1}=\mathbb{B}_f=\{\mathbb{B}_f(e)|e\in\gamma\},\ \ \mathbb{B}_{0}= \mathbb{B}^g_i=\{\mathbb{B}^g_i(e)|e\in\gamma\}$;
Besides, the effective action ${S}[\mathbb{B}^\ast, \mathbb{B}]$ is given by
\begin{equation}\label{actiontg2}
{S}[\mathbb{B}^\ast, \mathbb{B}]=\sum_{\imath=0}^{N}-\mathbf{i}\mathfrak{I}(\mathbb{B}_{\imath+1},\mathbb{B}_{\imath+1}^\ast; \mathbb{B}_\imath, \mathbb{B}_\imath^\ast)-\frac{\kappa}{a^2}\sum_{\imath=0}^{N}\Delta \tau[\mathbf{H}_{\imath+1,\imath}+\mathbf{i}\tilde{\varepsilon}_{\imath+1,\imath}(\frac{\Delta \tau}{\hbar})],
\end{equation}
where we define $\mathbf{H}_{\imath+1,\imath}\equiv \frac{\langle\tilde{\Psi}^t_{\mathbb{B}_{\imath+1}} |\hat{\mathbf{H}}|\tilde{\Psi}^t _{\mathbb{B}_{\imath}}\rangle }{\langle\tilde{\Psi}^t_{\mathbb{B}_{\imath+1}}|\tilde{\Psi}^t_{{\mathbb{B}}_{\imath}}\rangle }$, and $\mathfrak{I}(\mathbb{B}_{\imath+1},\mathbb{B}_{\imath+1}^\ast; \mathbb{B}_\imath, \mathbb{B}_\imath^\ast)$ is given by Eqs. \eqref{KIrelation} with its properties being given by \eqref{oneformCcontinuum}.
{Considering the continuous limit as usually done in the path integral formulation, we have} 
\begin{eqnarray}\label{AHH3}
{K}(\mathbb{B}_f,\mathbb{B}_i,T)&=&\lim_{N\to \infty} \int\prod_{\imath=1}^{N}\prod_{e}d\nu(\mathbb{B}^\ast_\imath(e),\mathbb{B}_\imath(e))\exp(\mathbf{i}{S}[\mathbb{B}^\ast, \mathbb{B}]/t)\\\nonumber
&=& \int d\mu[\mathbb{B}^\ast,\mathbb{B}]\exp(\mathbf{i}\mathcal{S}[\mathbb{B}^\ast,\mathbb{B}]/t),
\end{eqnarray}
where the path measure $d\mu[\mathbb{B}^\ast,\mathbb{B}]$ is given  by
\begin{equation}\label{pathmea1}
d\mu[\mathbb{B}^\ast,\mathbb{B}]=\lim_{N\to\infty} \prod_{\imath=1}^{N}\prod_{e}d\nu(\mathbb{B}^\ast_\imath(e),\mathbb{B}_\imath(e))
\end{equation}
and the action $\mathcal{S}[\mathbb{B}^\ast,\mathbb{B}]=\lim_{N\to\infty} {S}[\mathbb{B}^\ast, \mathbb{B}]$ reads
\begin{equation}
\mathcal{S}[\mathbb{B}^\ast,\mathbb{B}]=\int d\tau (-\mathbf{i})\mathfrak{I}[\mathbb{B}^\ast(\tau), \mathbb{B}(\tau),\dot{\mathbb{B}}^\ast(\tau), \dot{\mathbb{B}}(\tau)]-\frac{\kappa}{a^2}\int d\tau\mathbf{H}[\mathbb{B}^\ast(\tau),\mathbb{B}(\tau)],
\end{equation}
with $\mathbf{H}[\mathbb{B}^\ast(\tau),\mathbb{B}(\tau)]:={\langle\tilde{\Psi}^t_{\mathbb{B}(\tau)} |\hat{\mathbf{H}}|\tilde{\Psi}^t _{\mathbb{B}(\tau)}\rangle }$ and 
\begin{eqnarray}
&& \mathfrak{I}[\mathbb{B}^\ast(\tau), \mathbb{B}(\tau),\dot{\mathbb{B}}^\ast(\tau), \dot{\mathbb{B}}(\tau)]\\\nonumber
&:=&\lim_{\Delta\tau\to0}\frac{\mathfrak{K}(\mathbb{B}_{\imath+1},\mathbb{B}_{\imath+1}^\ast; \mathbb{B}_\imath, \mathbb{B}_\imath^\ast)}{\Delta\tau}+\frac{\mathbf{i}}{4}\frac{(\eta^{\imath+1}_e\xi^{\imath+1}_e-\eta^{\imath}_e\xi^{\imath}_e)}{\Delta\tau}\\\nonumber
&=&\sum_{e}\frac{1}{2}(Z_e\frac{d{W}_e}{d\tau}-W_e \frac{d{Z}_e}{d\tau})+\sum_{v}\frac{1}{\sqrt{2}}(\breve{Z}_v+\breve{W}_v)(\frac{\check{Z}_v \frac{d\check{W}_v}{d\tau}-\check{W}_v\frac{ d\check{Z}_v}{d\tau}}{1+\check{Z}_v \check{W}_v})\\\nonumber
&&+\sum_{v}\sum_{I=1}^{4}\frac{1}{2}({{Z}_{v,I} \frac{d{W}_{v,I}}{d\tau}-\frac{d{Z}_{v,I}}{d\tau} {W}_{v,I}})+\sum_{v}\frac{1}{2}(\breve{Z}_v\frac{d\breve{W}_v}{d\tau}-\breve{W}_v\frac{ d\breve{Z}_v}{d\tau})+\frac{\mathbf{i}}{4}\sum_{e}\frac{d(\eta_e\xi_e)}{d\tau}
\end{eqnarray}
derived by using Eq.\eqref{oneformBcontinuum}.

Now, let us consider the treatment of the gauge degrees of freedom in the action ${S}[\mathbb{B}^\ast, \mathbb{B}]$. 
We first have the variation with respect to $g=\{g_v\}_{v\in\gamma}$  acting on the initial state, which gives
  \begin{equation}\label{clousure}
  G_v(\mathbb{B}_i,\mathbb{B}_i^\ast)=0 \quad \Leftrightarrow \quad  \breve{Z}_{v,i}+ \breve{Z}^\ast_{v,i}=0,
 \end{equation}
 for all $v\in\gamma$. Moreover, we have the variation with respect to the pure gauge variables  $(\check{Z}^\imath_v,\check{W}^\imath_v,\breve{Z}^\imath_{v},\breve{W}^\imath_{v},Z^\imath_{v,4},W^\imath_{v,4})$.
Since the physical Hamiltonian operator $\hat{\mathbf{H}}$ is gauge invariant with respect to Gauss constraint,  it is reasonable to assume 
 \begin{equation}\label{con1}
\frac{\partial\mathbf{ H}_{\imath+1,\imath}}{\partial \check{Z}^{\imath}_v}=\frac{\partial \mathbf{ H}_{\imath+1,\imath}}{\partial \check{W}^{\imath+1}_v}=\frac{\partial \mathbf{H}_{\imath+1,\imath}}{\partial \breve{W}^{\imath+1}_{v}}=\frac{\partial \mathbf{H}_{\imath+1,\imath}}{\partial \breve{Z}^{\imath}_{v}}=\frac{\partial \mathbf{H}_{\imath+1,\imath}}{\partial W^{\imath+1}_{v,4}}=\frac{\partial \mathbf{H}_{\imath+1,\imath}}{\partial Z^{\imath}_{v,4}}=\mathcal{O}(\Delta\tau)
 \end{equation}
 on the full phase space, {which ensures that} there is no dynamics for the pure gauge variables $(\check{Z}^{\imath}_v,\check{W}^{\imath}_v,\breve{Z}^{\imath}_{v},\breve{W}^{\imath}_{v},Z^{\imath}_{v,4},W^{\imath}_{v,4})$ in the time continuum limit $\Delta\tau\to 0$. Then, 
 %by fixing the boundary data $\check{W}^\ast_{v,f}=\check{Z}_{v,i},\breve{W}_{v,f}^\ast=\breve{Z}_{v,i}$ and $W_{v,4,f}^\ast=Z_{v,4,i}$ for the  variables,
 one can  perform the integral  in \eqref{AHH3} over the  pure gauge variables $(\check{Z}^{\imath}_v,\check{W}^{\imath}_v,\breve{Z}^{\imath}_{v},\breve{W}^{\imath}_{v},Z^{\imath}_{v,4},W^{\imath}_{v,4})$  directly, which leads to 
 \begin{eqnarray}\label{AHHfinal}
{K}(\mathbb{B}_f,\mathbb{B}_i,T)
&=&\int d\tilde{\mu}[\mathbb{B}^\ast,\mathbb{B}]\exp(\mathbf{i}\tilde{\mathcal{S}}[\mathbb{B}^\ast,\mathbb{B}]/t)
\end{eqnarray}
 where we define $d\tilde{\mu}[\mathbb{B}^\ast,\mathbb{B}]:=\lim_{N\to \infty}\prod_{\imath=1}^{N}d\tilde{\nu}(\mathbb{B}_\imath^\ast,\mathbb{B}_\imath)$ with 
\begin{eqnarray}\label{mea5}
d\tilde{\nu}(\mathbb{B}^\ast,\mathbb{B})
:=\prod_{e}\frac{dZ_edW_e}{2t\pi\mathbf{i}}\prod_{v}\prod_{I=1}^{3}\frac{dZ_{v,I}dW_{v,I}}{2t\pi\mathbf{i}}(1+\mathcal{O}(\frac{t}{\eta})+\mathcal{O}(e^{-\frac{1}{t}}))
\end{eqnarray}
 and
the reduced action $\tilde{\mathcal{S}}[\mathbb{B}^\ast,\mathbb{B}]$ given by
\begin{eqnarray}
\tilde{\mathcal{S}}[\mathbb{B}^\ast,\mathbb{B}]&=&\int d\tau (-\mathbf{i})\tilde{\mathfrak{I}}[\mathbb{B}^\ast(\tau), \mathbb{B}(\tau),\dot{\mathbb{B}}^\ast(\tau), \dot{\mathbb{B}}(\tau)]-\frac{\kappa}{a^2}\int d\tau\mathbf{H}[\mathbb{B}^\ast(\tau),\mathbb{B}(\tau)]+\mathcal{J}(\mathbb{B}_f,\mathbb{B}_i^g)
\end{eqnarray}
with
\begin{eqnarray}
&&\tilde{ \mathfrak{I}}[\mathbb{B}^\ast(\tau), \mathbb{B}(\tau),\dot{\mathbb{B}}^\ast(\tau), \dot{\mathbb{B}}(\tau)]\\\nonumber
&:=&\sum_{e}\frac{1}{2}(Z_e\dot{W}_e-W_e \dot{Z}_e)+\sum_{v}\sum_{I=1}^{3}\frac{1}{2}({Z}_{v,I} \dot{W}_{v,I}-\dot{Z}_{v,I} {W}_{v,I})+\frac{\mathbf{i}}{4}\sum_{e}\frac{d(\eta_e\xi_e)}{d\tau}
\end{eqnarray}
and 
\begin{eqnarray}
&&\mathcal{J}(\mathbb{B}_f,\mathbb{B}_i^g)\\\nonumber
&:=&-\mathbf{i}t\ln\left(\langle\tilde{\Psi}_{\mathbb{B}_f}^t|g^{-1}|\tilde{\Psi}_{\mathbb{B}_i}^t\rangle\right)-\sum_{e}(Z_e^iW_e^f-\frac{1}{2}|Z^i_e|^2-\frac{1}{2}|W^f_e|^2)\\\nonumber
 &&-\sum_{v}\sum_{I=1}^{3}(Z_{v,I}^iW_{v,I}^f-\frac{1}{2}|Z^i_{v,I}|^2-\frac{1}{2}|W^f_{v,I}|^2).
\end{eqnarray}
Here $\mathcal{J}(\mathbb{B}_f,\mathbb{B}_i^g)$ is the boundary term for the pure gauge variables.

Now, 
%the EOMs are given by Eqs.\eqref{EEOM515} and \eqref{clousure} completely. Also, 
the variation of the action $\tilde{\mathcal{S}}[\mathbb{B}^\ast,\mathbb{B}]$ gives the following EOMs, 
  \begin{equation}\label{EEOM51}
   \mathbf{i}\dot{W}_e=-\frac{\kappa}{a^2}\frac{\partial \mathbf{H}}{\partial Z_e},\quad  \mathbf{i}\dot{Z}_e=\frac{\kappa}{a^2}\frac{\partial \mathbf{H}}{\partial W_e},\quad    \mathbf{i}\dot{W}_{v,I}=-\frac{\kappa}{a^2}\frac{\partial \mathbf{H}}{\partial Z_{v,I}},\quad  \mathbf{i}\dot{Z}_{v,I}=\frac{\kappa}{a^2}\frac{\partial \mathbf{H}}{\partial W_{v,I}}, \ I=1,2,3.
 \end{equation}
% which are just the time-continuum limit of Eqs.\eqref{EEOM515} .
{ The  EOMs \eqref{EEOM51} may } give a complex path for given boundary states labelled by $(\mathbb{B}_f,\mathbb{B}_i)$. 
 Thus, we need duplicate the reduced phase space $H_\gamma^+$ by relaxing the dependence  of $(W_e,{W}_{v,I})$ on $(Z_e,{Z}_{v,I})$ for $I=1,2,3$ \cite{PhysRevE.69.066204}. More precisely, we will assume that $(W_e,{W}_{v,I})$ are no longer the complex conjugate of $(Z_e,{Z}_{v,I})$.  One should notice that this duplication must preserve the initial and final data. {Then,} the  action $\tilde{\mathcal{S}}[\mathbb{B}^\ast,\mathbb{B}]$ in the duplicated  phase space $H_\gamma^+$ is given by
\begin{eqnarray}\label{Sfinal}
\tilde{\mathcal{S}}[\mathbb{B}^\ast,\mathbb{B}]&=&\int d\tau (-\mathbf{i})\tilde{\mathfrak{I}}[\mathbb{B}^\ast(\tau), \mathbb{B}(\tau),\dot{\mathbb{B}}^\ast(\tau), \dot{\mathbb{B}}(\tau)]-\frac{\kappa}{a^2}\int d\tau\mathbf{H}[\mathbb{B}^\ast(\tau),\mathbb{B}(\tau)]+\mathcal{J}(\mathbb{B}_f,\mathbb{B}_i^g)\\\nonumber
&&+\sum_{e}\left(\frac{1}{2}(Z_e^iW_e(0)+Z_e(T)W_e^f)-\frac{1}{2}(|Z^i_e|^2+|W^f_e|^2)\right)\\\nonumber
 &&+\sum_{v}\sum_{I=1}^{3}\left(\frac{1}{2}(Z_{v,I}^iW_{v,I}(0)+Z_{v,I}(T)W_{v,I}^f)-\frac{1}{2}(|Z^i_{v,I}|^2+|W^f_{v,I}|^2)\right)
\end{eqnarray}
with $\tilde{ \mathfrak{I}}[\mathbb{B}^\ast(\tau), \mathbb{B}(\tau),\dot{\mathbb{B}}^\ast(\tau), \dot{\mathbb{B}}(\tau)]$ being redefined as
\begin{eqnarray}\label{symterm555}
&&\tilde{ \mathfrak{I}}[\mathbb{B}^\ast(\tau), \mathbb{B}(\tau),\dot{\mathbb{B}}^\ast(\tau), \dot{\mathbb{B}}(\tau)]\\\nonumber
&:=&\sum_{e}\frac{1}{2}(Z_e\dot{W}_e-W_e \dot{Z}_e)+\sum_{v}\sum_{I=1}^{3}\frac{1}{2}({Z}_{v,I} \dot{W}_{v,I}-\dot{Z}_{v,I} {W}_{v,I})+\frac{\mathbf{i}}{4}\sum_{e}\frac{d(\eta_e\xi_e)}{d\tau},
\end{eqnarray}
where  $W_e(0)=W_e|_{\tau=0}$ is not the complex conjugate of $Z_e^i$, $Z_e(T)=Z_e|_{\tau=T}$ is not the complex conjugate of $W_e^f$, and likewise for $W_{v,I}(0)$, $Z_{v,I}(T)$.

{It is observed} that the measure \eqref{mea5}  and the symplectic term \eqref{symterm555} takes the  similar formulation as that for multiple-particle quantum system in QM. 
Hence, the quantum propagator \eqref{AHHfinal} with the action \eqref{Sfinal} can be computed based on the semi-classical approximation methods. Let us carry out this computation as follows.
First, let us define the discrete version of 
 $\tilde{\mathcal{S}}[\mathbb{B}^\ast,\mathbb{B}]$ as
\begin{eqnarray}
\tilde{{S}}[\mathbb{B}^\ast,\mathbb{B}]&:=&\sum_{\imath=0}^{N}-\mathbf{i}\tilde{I}(\mathbb{B}_{\imath+1},\mathbb{B}_{\imath+1}^\ast; \mathbb{B}_\imath, \mathbb{B}_\imath^\ast)-\frac{\kappa}{a^2}\sum_{\imath=0}^{N}\Delta \tau[\mathbf{H}_{\imath+1,\imath}+\mathbf{i}\tilde{\varepsilon}_{\imath+1,\imath}(\frac{\Delta \tau}{\hbar})]+\mathcal{J}(\mathbb{B}_f,\mathbb{B}_i^g),
\end{eqnarray}
where 
\begin{eqnarray}
&&\tilde{I}(\mathbb{B}_{\imath+1},\mathbb{B}_{\imath+1}^\ast; \mathbb{B}_\imath, \mathbb{B}_\imath^\ast)\\\nonumber
&:=&\sum_{e}\frac{1}{2}\left(Z^{\imath}_e({W}^{\imath+1}_e-{W}^{\imath}_e)-W^{\imath+1}_e ({Z}^{\imath+1}_e-{Z}^{\imath}_e)\right)\\\nonumber
&&+\sum_{v}\sum_{I=1}^{3}\frac{1}{2}\left({Z}^{\imath}_{v,I} ({W}^{\imath+1}_{v,I}-{W}^{\imath}_{v,I})-{W}^{\imath+1}_{v,I}({Z}^{\imath+1}_{v,I}-{Z}^{\imath}_{v,I}) \right)+\frac{\mathbf{i}}{4}\sum_{e}(\eta^{\imath+1}_e\xi^{\imath+1}_e-\eta^{\imath}_e\xi^{\imath}_e).
\end{eqnarray}
It is easy to verify that $\tilde{{S}}[\mathbb{B}^\ast,\mathbb{B}]$ satisfies $\tilde{\mathcal{S}}[\mathbb{B}^\ast, \mathbb{B}]=\lim_{N\to\infty }\tilde{S}[\mathbb{B}^\ast, \mathbb{B}]$. Also, the discrete version of the EOMs  \eqref{EEOM51} can be given by the variation of $\tilde{{S}}[\mathbb{B}^\ast,\mathbb{B}]$, which reads
\begin{eqnarray}\label{EEOMdiscrete}
  &&\mathbf{i}({W}^{\imath+1}_e-{W}^{\imath}_e)=-\Delta\tau\frac{\kappa}{a^2}\frac{\partial \mathbf{H}_{\imath+1,\imath}}{\partial Z^{\imath}_e},\quad   \mathbf{i}({W}^{\imath+1}_{v,I}-{W}^{\imath}_{v,I})=-\Delta\tau\frac{\kappa}{a^2}\frac{\partial \mathbf{H}_{\imath+1,\imath}}{\partial Z^{\imath}_{v,I}},\quad \imath=1,...,N,\\\nonumber 
  &&\mathbf{i}({Z}^{\imath+1}_e-{Z}^{\imath}_e)=\Delta\tau\frac{\kappa}{a^2}\frac{\partial \mathbf{H}_{\imath+1,\imath}}{\partial W^{\imath+1}_e},\quad  \mathbf{i}({Z}^{\imath+1}_{v,I}-{Z}^{\imath}_{v,I})=\Delta\tau\frac{\kappa}{a^2}\frac{\partial \mathbf{H}_{\imath+1,\imath}}{\partial W^{\imath+1}_{v,I}}, \quad  \imath=0,...,N-1,
\end{eqnarray}
 for $I=1,2,3,$ and $e\in\gamma,v\in\gamma$.
Then, the semiclassical approximation of the quantum propagator 
is given by expanding the action, for fixed boundary conditions and time interval, up to the second perturbative order around the complex trajectories determined by the EOMs  \eqref{EEOM51}, leading to
  \begin{equation}
  \tilde{\mathcal{S}}\approx  \tilde{\mathcal{S}}^{\text{c}}+\frac{1}{2}\delta^2  \tilde{\mathcal{S}}^{\text{c}},\quad \delta  \tilde{\mathcal{S}}^{\text{c}}=0.
 \end{equation}
Further, one has \cite{PhysRevLett.79.3323,PhysRevE.69.066204,MBaranger_2001,10.1063/1.3583996,10.1063/1.4936315}
 \begin{equation}\label{Ksc1}
{K}(\mathbb{B}_f,\mathbb{B}_i,T)\approx\sum_{\text{c.t.}}{K}_{\text{reduced}}(\mathbb{B}^\ast_f,\mathbb{B}_i,T)\int dg\exp(\mathbf{i}\tilde{\mathcal{S}}^{\text{c}}[\mathbb{B}_f^\ast, \mathbb{B}_i,T]/t),
 \end{equation}
where  the summation is over all complex trajectories and we define
\begin{eqnarray}\label{Kred1}
{K}_{\text{reduced}}(\mathbb{B}^\ast_f,\mathbb{B}_i,T)&\equiv&\int d\tilde{\mu}(\mathbb{B}^\ast,\mathbb{B})\exp(\frac{\mathbf{i}}{2t}\delta^2\tilde{\mathcal{S}}^{\text{c}}[\mathbb{B}^\ast, \mathbb{B}])\\\nonumber
&=&\lim_{N\to\infty }\int  \prod_{\imath=1}^Nd\tilde{\nu}(\mathbb{B}_\imath^\ast,\mathbb{B}_\imath)\exp(\frac{\mathbf{i}}{2t}\delta^2\tilde{S}^{\text{c}}[\mathbb{B}^\ast, \mathbb{B}])
%&=&\\\nonumber
%&=&\sqrt{\det\left(\mathbf{i}\frac{\partial \tilde{\mathcal{S}}^{\text{c}}[\mathbb{B}^\ast, \mathbb{B}]}{ \partial{\vec{W}}(T)\partial{\vec{Z}}(0) }\right)}\exp\left(\frac{\mathbf{i}t}{2\hbar}\int_{0}^T d\tau\left( \sum_{e}\frac{\partial ^2\mathbf{H}}{\partial Z_e\partial W_e}+\sum_{v}\sum_{I=1}^3\frac{\partial ^2\mathbf{H}}{\partial Z_{v,I}\partial W_{v,I}}\right)\right),
 \end{eqnarray}
 with $\delta^2\tilde{S}^{\text{c}}[\mathbb{B}^\ast, \mathbb{B}]$ being given by the expansion
   \begin{equation}
  \tilde{{S}}\approx  \tilde{{S}}^{\text{c}}+\frac{1}{2}\delta^2  \tilde{{S}}^{\text{c}},\quad \delta  \tilde{{S}}^{\text{c}}=0
 \end{equation}
 around the complex trajectories determined by the discrete EOMs\eqref{EEOMdiscrete}. 
To  perform the integral in ${K}_{\text{reduced}}(\mathbb{B}^\ast_f,\mathbb{B}_i,T)$, let us introduce the $2\mathcal{L} \times 1$ matrix  $\vec{M}:=(\vec{W},\vec{Z})$ with $\mathcal{L}:=N(|E(\gamma)|+3|V(\gamma)|)$, where the  $\mathcal{L}\times1$ matrix $\vec{W}$ is defined by $\vec{W}:=(...,(...,W^\imath_e,...,W^\imath_{v,I},...),...)|_{\imath=1,...,N}$  and the  $\mathcal{L}\times1$ matrix $\vec{Z}$ is defined by  $\vec{Z}:=(...,(...,Z^\imath_e,...,Z^\imath_{v,I},...),...)|_{\imath=1,...,N}$. Then, one can introduce the  $2\mathcal{L}\times2\mathcal{L}$  matrix $\mathcal{D}^{(\mathcal{L})}$  defined by  
   \begin{eqnarray}
\delta \vec{M}^{\text{T}}\mathcal{D}^{(\mathcal{L})}\delta\vec{M}:=-\frac{\mathbf{i}}{t}{\delta^2\tilde{S}^{\text{c}}[\mathbb{B}^\ast, \mathbb{B}]},
 \end{eqnarray}
 where $\delta \vec{M}:=\vec{M}-\vec{M}^{\text{c}}$, with $\vec{M}^{\text{c}}$ being the value of  $\vec{M}$ on  the complex trajectories determined by the discrete EOMs\eqref{EEOMdiscrete}, and  $\delta \vec{M}^{\text{T}}$ being the transposition of $\delta \vec{M}$.
  Now, the path integral in Eq.\eqref{Kred1} is just a Gaussian integral. {Thus} one has \cite{MBaranger_2001,10.1063/1.4936315,Torre:2005yy}
 \begin{eqnarray}\label{Kred2}
{K}_{\text{reduced}}(\mathbb{B}^\ast_f,\mathbb{B}_i,T)
&=&\lim_{N\to\infty }\int  \prod_{\imath=1}^Nd\tilde{\nu}(\mathbb{B}_\imath^\ast,\mathbb{B}_\imath)\exp(\frac{\mathbf{i}}{2t}\delta^2\tilde{S}^{\text{c}}[\mathbb{B}^\ast, \mathbb{B}])\\\nonumber
&=&\lim_{N\to\infty }\int  \prod_{\imath=1}^Nd\tilde{\nu}(\mathbb{B}_\imath^\ast,\mathbb{B}_\imath)\exp(-\frac{1}{2}\delta \vec{M}^{\text{T}}\mathcal{D}^{(\mathcal{L})}\delta\vec{M})\\\nonumber
&=&\lim_{N\to\infty }\frac{1}{(\mathbf{i}t)^{\mathcal{L}}\sqrt{\det(\mathcal{D}^{(\mathcal{L})})}}.
%&=&\sqrt{\det\left(\mathbf{i}\frac{\partial \tilde{\mathcal{S}}^{\text{c}}[\mathbb{B}^\ast, \mathbb{B}]}{ \partial{\vec{W}}(T)\partial{\vec{Z}}(0) }\right)}\exp\left(\frac{\mathbf{i}t}{2\hbar}\int_{0}^T d\tau\left( \sum_{e}\frac{\partial ^2\mathbf{H}}{\partial Z_e\partial W_e}+\sum_{v}\sum_{I=1}^3\frac{\partial ^2\mathbf{H}}{\partial Z_{v,I}\partial W_{v,I}}\right)\right),
 \end{eqnarray}
{Due to}  that the symplectic term  in $\tilde{\mathcal{S}}^{\text{c}}[\mathbb{B}^\ast, \mathbb{B}]=\lim_{N\to\infty }\tilde{S}^{\text{c}}[\mathbb{B}^\ast, \mathbb{B}]$ is identical to that for the multiple-particle quantum system in QM, ${K}_{\text{reduced}}(\mathbb{B}^\ast_f,\mathbb{B}_i,T)$ can be further simplified by following the relevant results in Refs. \cite{MBaranger_2001,10.1063/1.3583996,10.1063/1.4936315}, which leads to
  \begin{eqnarray}\label{Kred3}
{K}_{\text{reduced}}(\mathbb{B}^\ast_f,\mathbb{B}_i,T)
&=&\lim_{N\to\infty }\frac{1}{(\mathbf{i}t)^{\mathcal{L}}\sqrt{\det(\mathcal{D}^{(\mathcal{L})})}}\\\nonumber
&=&\sqrt{\det\left(\frac{\partial {{W}}(0)}{ \partial{{W}}(T) }\right)}\exp\left(\frac{\mathbf{i}\kappa}{2a^2}\int_{0}^T d\tau\left( \sum_{e}\frac{\partial ^2\mathbf{H}}{\partial Z_e\partial W_e}+\sum_{v}\sum_{I=1}^3\frac{\partial ^2\mathbf{H}}{\partial Z_{v,I}\partial W_{v,I}}\right)\right),
 \end{eqnarray}
 where ${W}(T):=(...,W_e,...,W_{v,I},...)|_{\tau=T}=(...,W^f_e,...,W^f_{v,I},...)$ and ${W}(0):=(...,W_e,...,W_{v,I},...)|_{\tau=0}$  for $e\in E(\gamma)$, $v\in V(\gamma)$, and $I\in \{1,2,3\}$. By denoting 
  \begin{eqnarray}
\mathcal{I}_{\text{reduced}}
&\equiv&\frac{t\kappa}{2a^2}\int_{0}^T d\tau\left( \sum_{e}\frac{\partial ^2\mathbf{H}}{\partial Z_e\partial W_e}+\sum_{v}\sum_{I=1}^3\frac{\partial ^2\mathbf{H}}{\partial Z_{v,I}\partial W_{v,I}}\right),
 \end{eqnarray}
 the final result of the semi-classical approximation of the propagator is given by
  \begin{equation}\label{Kscfinal}
{K}(\mathbb{B}_f,\mathbb{B}_i,T)\approx\sum_{\text{c.t.}}\sqrt{\det\left(\frac{\partial {{W}}(0)}{ \partial{{W}}(T) }\right)}\exp\left(\mathbf{i}  \mathcal{I}_{\text{reduced}} /t \right)\int dg\exp(\mathbf{i}\tilde{\mathcal{S}}^{\text{c}}[\mathbb{B}_f^\ast, \mathbb{B}_i,T]/t).
 \end{equation}

It is worth to {having} some discussions on the above result, which are listed as follows. 
\begin{enumerate}
\item The the  $2\mathcal{L}\times2\mathcal{L}$  matrix $-\mathcal{D}^{(\mathcal{L})}$  is called the Hessian matrix in Ref.\cite{Han:2020chr}. The determinant of this matrix can be calculated in several methods \cite{AVRAMIDI20023,Dunne_2008,doi:10.1142/7305}. In above calculations,  the  computation skill for the same issue in the multiple-particle quantum system in QM is applied \cite{MBaranger_2001,10.1063/1.3583996,10.1063/1.4936315}, since the symplectic term  in $\tilde{S}^{\text{c}}[\mathbb{B}^\ast, \mathbb{B}]$ is identical that in QM.

\item One should notice that the semi-classical propagator ${K}(\mathbb{B}_f,\mathbb{B}_i,T)$ is not normalized, since the 
boundary states are not normalized for the gauge invariant formulation.  The normalized version of ${K}(\mathbb{B}_f,\mathbb{B}_i,T)$ reads
\begin{equation}
{K}_{\text{norm}}(\mathbb{B}_f,\mathbb{B}_i,T)=\frac{{K}(\mathbb{B}_f,\mathbb{B}_i,T)}{\sqrt{\int dg'\langle\tilde{\Psi}_{\mathbb{B}_i}^t|g'^{-1}|\tilde{\Psi}_{\mathbb{B}_i}^t\rangle}\sqrt{\int dg''\langle\tilde{\Psi}_{\mathbb{B}_f}^t|g''^{-1}|\tilde{\Psi}_{\mathbb{B}_f}^t\rangle}}.
 \end{equation}
Note that ${K}(\mathbb{B}_f,\mathbb{B}_i,T)$ and its normalization factor contain the group averaging operation. In fact, the group averaging only acts on the intertwiners  in the coherent state, which leads to the coherent intertwiners. Moreover, the coherent intertwiners can be  expanded in the orthonormal and gauge-invariant intertwiner basis, and the expansion coefficients have been calculated in Ref. \cite{Long:2024lbd}.
\item The factor $\det\left(\frac{\partial {{W}}(0)}{ \partial{{W}}(T) }\right)$ can be calculated based on the trajectory equation given by solving the complexified EOMs  \eqref{EEOM51}. Specifically, the trajectory equations take the formulation 
  \begin{eqnarray}\label{trajecequation}
W_e(\tau)&=&W_e(\tau, ...,W^f_e,...,W^f_{v,I},...,...,Z^i_e,...,Z^i_{v,I},...),\\\nonumber
W_{v,I}(\tau)&=&W_{v,I}(\tau, ...,W^f_e,...,W^f_{v,I},...,...,Z^i_e,...,Z^i_{v,I},...),\\\nonumber
 Z_e(\tau)&=&Z_e(\tau, ...,W^f_e,...,W^f_{v,I},...,...,Z^i_e,...,Z^i_{v,I},...),\\\nonumber
Z_{v,I}(\tau)&=&Z_{v,I}(\tau, ...,W^f_e,...,W^f_{v,I},...,...,Z^i_e,...,Z^i_{v,I},...).
 \end{eqnarray}
Note that ${W}(T)=(...,W^f_e,...,W^f_{v,I},...)$,  it is obviously that $\det\left(\frac{\partial {{W}}(0)}{ \partial{{W}}(T) }\right)$ can be given by the variation of  the trajectory equations \eqref{trajecequation}.
\end{enumerate}

Finally, one can conclude that the semiclassical approximation of the quantum  propagator ${K}(\mathbb{B}_f,\mathbb{B}_i,T)$ is determined by the complex  trajectories, which can be given by solving the complexified EOMs  \eqref{EEOM51}. Besides, for each complex  trajectory, the final semi-classical propagator ${K}(\mathbb{B}_f,\mathbb{B}_i,T)$ is composed by two factors $\exp(\mathbf{i}\tilde{\mathcal{S}}^{\text{c}}[\mathbb{B}_f^\ast, \mathbb{B}_i,T]/t)$ and ${K}_{\text{reduced}}(\mathbb{B}^\ast_f,\mathbb{B}_i,T)$ , which  come from the zeroth and second order contributions of the complex trajectory respectively.

\section{Conclusion and Discussion}\label{sec:five}
A new formulation of coherent state path integral is established in this work. Specifically,  we introduce a new set of  geometric variables to parametrize the holonomy-flux phase space. Remarkably, we prove that this set of new geometric variables forms a rather simple Poisson algebra in the gauge invariant phase space. This result ensures that the  path measure and the symplectic form for the coherent state path-integral for twisted geometry takes a similar form as that for multiple-particle quantum system in QM. This allows us to compute this path-integral on twisted geometry analytically by the semiclassical approximation. The result shows that the semi-classical propagator is determined by the complex trajectories  purely, with the complex trajectories is given by solving the complexified EOMs.

Additionally, there are several other results worth discussing.
First, the  new geometric variables forms a simple Poisson algebra in the gauge invariant phase space, which may suggest a new quantization scheme of the twisted geometry, i.e., the Bohr-Sommerfeld quantization \cite{PhysRevLett.107.011301,Bianchi:2012wb}. 
Second, for fixed boundary data, one can calculate the propagator by the semi-classical approximation as shown in this article and by spin-foam models respectively; The comparison of the results given by these two path integral methods in LQG would be an interesting topic in future researches.
Third, for each complex trajectory, the final semi-classical propagator ${K}(\mathbb{B}_f,\mathbb{B}_i,T)$ contains two factors $\exp(\mathbf{i}\tilde{\mathcal{S}}^{\text{c}}[\mathbb{B}_f^\ast, \mathbb{B}_i,T]/t)$ and ${K}_{\text{reduced}}(\mathbb{B}^\ast_f,\mathbb{B}_i,T)$ , which  come from the zeroth-order and second-order contributions of the complex trajectory respectively. Comparing to the Euclidean path-integral of gravity which only contains the zeroth order contribution of the special Euclidean trajectory \cite{PhysRevD.18.1747,doi:10.1142/1301}, our result provides a more general framework to explore the quantum aspects of gravity, e.g., the thermodynamics of spacetime. 
Fourth, it is expected to extend this effective dynamics based on twisted geometry coherent path-integral for $SU(2)$ LQG in the (1+3)-dimensions to more general gauge field theory. In fact, the twisted geometry coherent state for general $SO(D+1)$ gauge field theory has been constructed and studied in several previous works  \cite{Long:2021lmd,Long:2022cex}. It is interesting to extend the construction in this article  to  $SO(D+1)$ LQG in arbitrary (1+D)-dimensions \cite{Bodendorfer:Ha,Bodendorfer:Qu,long2020operators,Long:2020agv,Long:2022thb}, to explore the dynamics of quantum geometry with extra dimensions.

\section*{Acknowledgments}

This work is supported by the project funded by  the National Natural Science Foundation of China (NSFC) with Grants No. 12405062 and No.12275022. G. L. is supported by the Fundamental Research Funds for the Central Universities with Grants No.21624340, and the Science and Technology Planning Project of Guangzhou with Grants No. 2024A04J4030.  H. L. is supported by research grants provided by the Blaumann Foundation. 

\bibliographystyle{unsrt}

\bibliography{ref}

\begin{thebibliography}{100}

\bibitem{first30years}
Abhay Ashtekar and Jorge Pulliny.
\newblock {\em Loop quantum gravity: The first 30 years}.
\newblock World Scientific Publishing Co. Pte Ltd, Singapore, mar 2017.

\bibitem{Ashtekar2012Background}
Abhay Ashtekar and Jerzy Lewandowski.
\newblock Background independent quantum gravity: a status report.
\newblock {\em Classical and Quantum Gravity}, 21(15):R53--R152, 2012.

\bibitem{RovelliBook2}
Carlo Rovelli and Francesca Vidotto.
\newblock {\em {Covariant Loop Quantum Gravity: An Elementary Introduction to
  Quantum Gravity and Spinfoam Theory}}.
\newblock Cambridge University Press, 2014.

\bibitem{Han2005FUNDAMENTAL}
Muxin Han, M.~A. Yongge, and Weiming Huang.
\newblock Fundamental structure of loop quantum gravity.
\newblock {\em International Journal of Modern Physics D}, 16(09):1397--1474,
  2005.

\bibitem{thiemann2007modern}
Thomas Thiemann.
\newblock {\em Modern canonical quantum general relativity}.
\newblock Cambridge University Press, 2007.

\bibitem{rovelli2007quantum}
Carlo Rovelli.
\newblock {\em Quantum gravity}.
\newblock Cambridge university press, 2007.

\bibitem{Zhang:2021qul}
Cong Zhang, Shicong Song, and Muxin Han.
\newblock {First-Order Quantum Correction in Coherent State Expectation Value
  of Loop-Quantum-Gravity Hamiltonian}.
\newblock {\em Phys. Rev. D}, 105:064008, 2022.

\bibitem{Long:2021izw}
Gaoping Long and Yongge Ma.
\newblock {Effective dynamics of weak coupling loop quantum gravity}.
\newblock {\em Phys. Rev. D}, 105(4):044043, 2022.

\bibitem{Zhang:2022vsl}
Cong Zhang, Hongguang Liu, and Muxin Han.
\newblock {Fermions in loop quantum gravity and resolution of doubling
  problem}.
\newblock {\em Class. Quant. Grav.}, 40(20):205022, 2023.

\bibitem{Liegener:2019jhj}
Klaus Liegener and Ernst-Albrecht Zwicknagel.
\newblock {Expectation values of coherent states for $SU(2)$ Lattice Gauge
  Theories}.
\newblock {\em JHEP}, 02:024, 2020.

\bibitem{Han:2024rqb}
Muxin Han, Dongxue Qu, and Cong Zhang.
\newblock {Spin foam amplitude of the black-to-white hole transition}.
\newblock 4 2024.

\bibitem{Han:2024ydv}
Muxin Han, Hongguang Liu, Dongxue Qu, Francesca Vidotto, and Cong Zhang.
\newblock {Cosmological Dynamics from Covariant Loop Quantum Gravity with
  Scalar Matter}.
\newblock 2 2024.

\bibitem{Assanioussi:2015gka}
Mehdi Assanioussi, Jerzy Lewandowski, and Ilkka Makinen.
\newblock {New scalar constraint operator for loop quantum gravity}.
\newblock {\em Phys. Rev.}, D92(4):044042, 2015.

\bibitem{Ashtekar:1996eg}
Abhay Ashtekar and Jerzy Lewandowski.
\newblock {Quantum theory of geometry. 1: Area operators}.
\newblock {\em Class. Quant. Grav.}, 14:A55--A82, 1997.

\bibitem{Ashtekar:1997fb}
Abhay Ashtekar and Jerzy Lewandowski.
\newblock {Quantum theory of geometry. 2. Volume operators}.
\newblock {\em Adv. Theor. Math. Phys.}, 1:388--429, 1998.

\bibitem{Bianchi:2008es}
Eugenio Bianchi.
\newblock {The Length operator in Loop Quantum Gravity}.
\newblock {\em Nucl. Phys. B}, 807:591--624, 2009.

\bibitem{Ma:2010fy}
Yongge Ma, Chopin Soo, and Jinsong Yang.
\newblock {New length operator for loop quantum gravity}.
\newblock {\em Phys. Rev. D}, 81:124026, 2010.

\bibitem{Giesel_2006Consistencycheck}
K~Giesel and T~Thiemann.
\newblock Consistency check on volume and triad operator quantization in loop
  quantum gravity: I.
\newblock {\em Classical and Quantum Gravity}, 23(18):5667, aug 2006.

\bibitem{Yang_2019Consistencycheck}
Jinsong Yang and Yongge Ma.
\newblock Consistency check on the fundamental and alternative flux operators
  in loop quantum gravity *.
\newblock {\em Chinese Physics C}, 43(10):103106, oct 2019.

\bibitem{ROVELLI1995593}
Carlo Rovelli and Lee Smolin.
\newblock Discreteness of area and volume in quantum gravity.
\newblock {\em Nuclear Physics B}, 442(3):593--619, 1995.

\bibitem{QoperatorPhysRevD.62.104021}
Yongge Ma and Yi~Ling.
\newblock $\mathrm{Q\ifmmode \hat{}\else \^{}\fi{}}$ operator for canonical
  quantum gravity.
\newblock {\em Phys. Rev. D}, 62:104021, Oct 2000.

\bibitem{volumePhysRevD.94.044003}
Jinsong Yang and Yongge Ma.
\newblock New volume and inverse volume operators for loop quantum gravity.
\newblock {\em Phys. Rev. D}, 94:044003, Aug 2016.

\bibitem{long2020operators}
Gaoping Long and Yongge Ma.
\newblock {General geometric operators in all dimensional loop quantum
  gravity}.
\newblock {\em Phys. Rev. D}, 101(8):084032, 2020.

\bibitem{Ashtekar:1997yu}
A.~Ashtekar, J.~Baez, A.~Corichi, and Kirill Krasnov.
\newblock {Quantum geometry and black hole entropy}.
\newblock {\em Phys. Rev. Lett.}, 80:904--907, 1998.

\bibitem{Ashtekar:2000eq}
A.~Ashtekar, John~C. Baez, and Kirill Krasnov.
\newblock {Quantum geometry of isolated horizons and black hole entropy}.
\newblock {\em Adv. Theor. Math. Phys.}, 4:1--94, 2000.

\bibitem{Long:2024lbd}
Gaoping Long, Qian Chen, and Jinsong Yang.
\newblock {Entanglement entropy of coherent intertwiner in loop quantum
  gravity}.
\newblock {\em Phys. Rev. D}, 110(6):064017, 2024.

\bibitem{Song:2020arr}
Shupeng Song, Haida Li, Yongge Ma, and Cong Zhang.
\newblock {Entropy of black holes with arbitrary shapes in loop quantum
  gravity}.
\newblock {\em Sci. China Phys. Mech. Astron.}, 64(12):120411, 2021.

\bibitem{Kaul:2000kf}
Romesh~K. Kaul and Parthasarathi Majumdar.
\newblock {Logarithmic correction to the Bekenstein-Hawking entropy}.
\newblock {\em Phys. Rev. Lett.}, 84:5255--5257, 2000.

\bibitem{Ghosh:2011fc}
Amit Ghosh and Alejandro Perez.
\newblock {Black hole entropy and isolated horizons thermodynamics}.
\newblock {\em Phys. Rev. Lett.}, 107:241301, 2011.
\newblock [Erratum: Phys.Rev.Lett. 108, 169901 (2012)].

\bibitem{Basu:2009cw}
Rudranil Basu, Romesh~K. Kaul, and Parthasarathi Majumdar.
\newblock {Entropy of Isolated Horizons revisited}.
\newblock {\em Phys. Rev. D}, 82:024007, 2010.

\bibitem{Engle:2010kt}
Jonathan Engle, Karim Noui, Alejandro Perez, and Daniele Pranzetti.
\newblock {Black hole entropy from an SU(2)-invariant formulation of Type I
  isolated horizons}.
\newblock {\em Phys. Rev. D}, 82:044050, 2010.

\bibitem{Song:2022zit}
Shupeng Song, Gaoping Long, Cong Zhang, and Xiangdong Zhang.
\newblock {Thermodynamics of isolated horizons in loop quantum gravity}.
\newblock {\em Phys. Rev. D}, 106(12):126007, 2022.

\bibitem{Donnelly:2008vx}
William Donnelly.
\newblock {Entanglement entropy in loop quantum gravity}.
\newblock {\em Phys. Rev. D}, 77:104006, 2008.

\bibitem{Perez:2014ura}
Alejandro Perez.
\newblock {Statistical and entanglement entropy for black holes in quantum
  geometry}.
\newblock {\em Phys. Rev. D}, 90(8):084015, 2014.
\newblock [Addendum: Phys.Rev.D 90, 089907 (2014)].

\bibitem{Dasgupta:2005yu}
Arundhati Dasgupta.
\newblock {Semi-classical quantisation of space-times with apparent horizons}.
\newblock {\em Class. Quant. Grav.}, 23:635--672, 2006.

\bibitem{Ashtekar:2003hd}
Abhay Ashtekar, Martin Bojowald, and Jerzy Lewandowski.
\newblock {Mathematical structure of loop quantum cosmology}.
\newblock {\em Adv. Theor. Math. Phys.}, 7(2):233--268, 2003.

\bibitem{Ashtekar:2011ni}
Abhay Ashtekar and Parampreet Singh.
\newblock {Loop Quantum Cosmology: A Status Report}.
\newblock {\em Class. Quant. Grav.}, 28:213001, 2011.

\bibitem{Long:2020oma}
Gaoping Long, Yunlong Liu, and Xiangdong Zhang.
\newblock {Energy conditions in the new model of loop quantum cosmology}.
\newblock {\em Chin. Phys. C}, 45(11):115102, 2021.

\bibitem{Zhang:2021zfp}
Xiangdong Zhang, Gaoping Long, and Yongge Ma.
\newblock {Loop quantum gravity and cosmological constant}.
\newblock {\em Phys. Lett. B}, 823:136770, 2021.

\bibitem{Bojowald:2001xe}
Martin Bojowald.
\newblock {Absence of singularity in loop quantum cosmology}.
\newblock {\em Phys. Rev. Lett.}, 86:5227--5230, 2001.

\bibitem{Ashtekar:2006rx}
Abhay Ashtekar, Tomasz Pawlowski, and Parampreet Singh.
\newblock {Quantum nature of the big bang}.
\newblock {\em Phys. Rev. Lett.}, 96:141301, 2006.

\bibitem{Ashtekar:2006wn}
Abhay Ashtekar, Tomasz Pawlowski, and Parampreet Singh.
\newblock {Quantum Nature of the Big Bang: Improved dynamics}.
\newblock {\em Phys. Rev. D}, 74:084003, 2006.

\bibitem{Ashtekar:2005qt}
Abhay Ashtekar and Martin Bojowald.
\newblock {Quantum geometry and the Schwarzschild singularity}.
\newblock {\em Class. Quant. Grav.}, 23:391--411, 2006.

\bibitem{Modesto:2005zm}
Leonardo Modesto.
\newblock {Loop quantum black hole}.
\newblock {\em Class. Quant. Grav.}, 23:5587--5602, 2006.

\bibitem{Boehmer:2007ket}
Christian~G. Boehmer and Kevin Vandersloot.
\newblock {Loop Quantum Dynamics of the Schwarzschild Interior}.
\newblock {\em Phys. Rev. D}, 76:104030, 2007.

\bibitem{Chiou:2012pg}
Dah-Wei Chiou, Wei-Tou Ni, and Alf Tang.
\newblock {Loop quantization of spherically symmetric midisuperspaces and loop
  quantum geometry of the maximally extended Schwarzschild spacetime}.
\newblock 12 2012.

\bibitem{Gambini:2013hna}
Rodolfo Gambini, Javier Olmedo, and Jorge Pullin.
\newblock {Quantum black holes in Loop Quantum Gravity}.
\newblock {\em Class. Quant. Grav.}, 31:095009, 2014.

\bibitem{Corichi:2015xia}
Alejandro Corichi and Parampreet Singh.
\newblock {Loop quantization of the Schwarzschild interior revisited}.
\newblock {\em Class. Quant. Grav.}, 33(5):055006, 2016.

\bibitem{Dadhich:2015ora}
Naresh Dadhich, Anton Joe, and Parampreet Singh.
\newblock {Emergence of the product of constant curvature spaces in loop
  quantum cosmology}.
\newblock {\em Class. Quant. Grav.}, 32(18):185006, 2015.

\bibitem{Olmedo:2017lvt}
Javier Olmedo, Sahil Saini, and Parampreet Singh.
\newblock {From black holes to white holes: a quantum gravitational, symmetric
  bounce}.
\newblock {\em Class. Quant. Grav.}, 34(22):225011, 2017.

\bibitem{Ashtekar:2018lag}
Abhay Ashtekar, Javier Olmedo, and Parampreet Singh.
\newblock {Quantum Transfiguration of Kruskal Black Holes}.
\newblock {\em Phys. Rev. Lett.}, 121(24):241301, 2018.

\bibitem{BenAchour:2018khr}
Jibril Ben~Achour, Fr{\'e}d{\'e}ric Lamy, Hongguang Liu, and Karim Noui.
\newblock {Polymer Schwarzschild black hole: An effective metric}.
\newblock {\em EPL}, 123(2):20006, 2018.

\bibitem{Han:2020uhb}
Muxin Han and Hongguang Liu.
\newblock {Improved effective dynamics of loop-quantum-gravity black hole and
  Nariai limit}.
\newblock {\em Class. Quant. Grav.}, 39(3):035011, 2022.

\bibitem{Kelly:2020lec}
Jarod~George Kelly, Robert Santacruz, and Edward Wilson-Ewing.
\newblock {Black hole collapse and bounce in effective loop quantum gravity}.
\newblock 6 2020.

\bibitem{Han:2022rsx}
Muxin Han and Hongguang Liu.
\newblock {Covariant ${\bar{\mu}}$-scheme effective dynamics, mimetic gravity,
  and non-singular black holes: Applications to spherical symmetric quantum
  gravity and CGHS model}.
\newblock 12 2022.

\bibitem{Giesel:2023hys}
Kristina Giesel, Hongguang Liu, Parampreet Singh, and Stefan~Andreas Weigl.
\newblock {Generalized analysis of a dust collapse in effective loop quantum
  gravity: fate of shocks and covariance}.
\newblock 8 2023.

\bibitem{Ashtekar:2023cod}
Abhay Ashtekar, Javier Olmedo, and Parampreet Singh.
\newblock {Regular black holes from Loop Quantum Gravity}.
\newblock 1 2023.

\bibitem{PhysRevLett.102.051301}
You Ding, Yongge Ma, and Jinsong Yang.
\newblock Effective scenario of loop quantum cosmology.
\newblock {\em Phys. Rev. Lett.}, 102:051301, Feb 2009.

\bibitem{Han_2020}
Muxin Han and Hongguang Liu.
\newblock Effective dynamics from coherent state path integral of full loop
  quantum gravity.
\newblock {\em Physical Review D}, 101(4), Feb 2020.

\bibitem{Han_2020semiclassical}
Muxin Han and Hongguang Liu.
\newblock Semiclassical limit of new path integral formulation from reduced
  phase space loop quantum gravity.
\newblock {\em Physical Review D}, 102(2), Jul 2020.

\bibitem{Qin:2011hx}
Li~Qin and Yongge Ma.
\newblock {Coherent State Functional Integrals in Quantum Cosmology}.
\newblock {\em Phys. Rev. D}, 85:063515, 2012.

\bibitem{Qin:2012xh}
Li~Qin and Yongge Ma.
\newblock {Coherent State Functional Integral in Loop Quantum Cosmology:
  Alternative Dynamics}.
\newblock {\em Mod. Phys. Lett. A}, 27:1250078, 2012.

\bibitem{Han:2021cwb}
Muxin Han and Hongguang Liu.
\newblock {Loop quantum gravity on dynamical lattice and improved cosmological
  effective dynamics with inflaton}.
\newblock {\em Phys. Rev. D}, 104(2):024011, 2021.

\bibitem{Han:2020iwk}
Muxin Han, Haida Li, and Hongguang Liu.
\newblock {Manifestly gauge-invariant cosmological perturbation theory from
  full loop quantum gravity}.
\newblock {\em Phys. Rev. D}, 102(12):124002, 2020.

\bibitem{1994The}
B.~Hall.
\newblock The segal-bargmann "coherent state" transform for compact lie groups.
\newblock {\em Journal of Functional Analysis}, 122(1):103--151, 1994.

\bibitem{ThiemannComplexifierCoherentStates}
T~Thiemann.
\newblock {Complexifier coherent states for quantum general relativity}.
\newblock {\em Class. Quantum Gravity}, 23:2063--2117, mar 2006.

\bibitem{Thomas2001Gauge}
Thomas Thiemann.
\newblock Gauge field theory coherent states (gcs): I. general properties.
\newblock {\em Classical and Quantum Gravity}, 18(11), 2001.

\bibitem{2001Gauge}
T.~Thiemann and O.~Winkler.
\newblock Gauge field theory coherent states (gcs): Ii. peakedness properties.
\newblock {\em Classical and Quantum Gravity}, 18(14):2561--2636, 2001.

\bibitem{2000Gauge}
T.~Thiemann and O.~Winkler.
\newblock Gauge field theory coherent states (gcs) : Iii. ehrenfest theorems.
\newblock {\em classical and quantum gravity}, 18(21):2561--2636, 2000.

\bibitem{Rovelli_2006}
Carlo Rovelli.
\newblock Graviton propagator from background-independent quantum gravity.
\newblock {\em Physical Review Letters}, 97(15), Oct 2006.

\bibitem{Bianchi_2009}
Eugenio Bianchi, Elena Magliaro, and Claudio Perini.
\newblock Lqg propagator from the new spin foams.
\newblock {\em Nuclear Physics B}, 822(1-2):245–269, Nov 2009.

\bibitem{Bianchi:2009ky}
Eugenio Bianchi, Elena Magliaro, and Claudio Perini.
\newblock {Coherent spin-networks}.
\newblock {\em Phys. Rev. D}, 82:024012, 2010.

\bibitem{Calcinari_2020}
Andrea Calcinari, Laurent Freidel, Etera Livine, and Simone Speziale.
\newblock Twisted geometries coherent states for loop quantum gravity.
\newblock {\em Classical and Quantum Gravity}, 38(2):025004, Dec 2020.

\bibitem{Long:2020euh}
Gaoping Long and Norbert Bodendorfer.
\newblock {Perelomov-type coherent states of SO($D+1$) in all-dimensional loop
  quantum gravity}.
\newblock {\em Phys. Rev. D}, 102(12):126004, 2020.

\bibitem{long2019coherent}
Gaoping Long, Chun-Yen Lin, and Yongge Ma.
\newblock Coherent intertwiner solution of simplicity constraint in all
  dimensional loop quantum gravity.
\newblock {\em Physical Review D}, 100(6):064065, 2019.

\bibitem{Livine:2007Nsfv}
Etera~R Livine and Simone Speziale.
\newblock New spinfoam vertex for quantum gravity.
\newblock {\em Physical Review D}, 76(8):084028, 2007.

\bibitem{PhysRevD.104.046014}
Gaoping Long, Cong Zhang, and Xiangdong Zhang.
\newblock Superposition type coherent states in all dimensional loop quantum
  gravity.
\newblock {\em Phys. Rev. D}, 104:046014, Aug 2021.

\bibitem{Long:2021lmd}
Gaoping Long, Xiangdong Zhang, and Cong Zhang.
\newblock {Twisted geometry coherent states in all dimensional loop quantum
  gravity: Construction and peakedness properties}.
\newblock {\em Phys. Rev. D}, 105(6):066021, 2022.

\bibitem{Long:2022cex}
Gaoping Long.
\newblock {Twisted geometry coherent states in all dimensional loop quantum
  gravity. II. Ehrenfest property}.
\newblock {\em Phys. Rev. D}, 106(6):066021, 2022.

\bibitem{Brown:1994py}
J.~David Brown and Karel~V. Kuchar.
\newblock {Dust as a standard of space and time in canonical quantum gravity}.
\newblock {\em Phys. Rev. D}, 51:5600--5629, 1995.

\bibitem{PhysRevD.43.419}
Karel~V. Kucha and Charles~G. Torre.
\newblock Gaussian reference fluid and interpretation of quantum
  geometrodynamics.
\newblock {\em Phys. Rev. D}, 43:419--441, Jan 1991.

\bibitem{Domagala:2010bm}
Marcin Domagala, Kristina Giesel, Wojciech Kaminski, and Jerzy Lewandowski.
\newblock {Gravity quantized: Loop Quantum Gravity with a Scalar Field}.
\newblock {\em Phys. Rev. D}, 82:104038, 2010.

\bibitem{PhysRevD.82.084040}
Laurent Freidel and Simone Speziale.
\newblock Twisted geometries: A geometric parametrization of su(2) phase space.
\newblock {\em Phys. Rev. D}, 82:084040, Oct 2010.

\bibitem{PhysRevD.103.086016}
Gaoping Long and Chun-Yen Lin.
\newblock Geometric parametrization of $\textsc{SO(D+1)}$ phase space of all
  dimensional loop quantum gravity.
\newblock {\em Phys. Rev. D}, 103:086016, Apr 2021.

\bibitem{Rovelli:2010km}
Carlo Rovelli and Simone Speziale.
\newblock {On the geometry of loop quantum gravity on a graph}.
\newblock {\em Phys. Rev.}, D82:044018, 2010.

\bibitem{PhysRevD.87.024038}
Hal~M. Haggard, Carlo Rovelli, Wolfgang Wieland, and Francesca Vidotto.
\newblock Spin connection of twisted geometry.
\newblock {\em Phys. Rev. D}, 87:024038, Jan 2013.

\bibitem{Bianchi:2010gc}
Eugenio Bianchi, Pietro Dona, and Simone Speziale.
\newblock {Polyhedra in loop quantum gravity}.
\newblock {\em Phys. Rev. D}, 83:044035, 2011.

\bibitem{PhysRevLett.107.011301}
Eugenio Bianchi and Hal~M. Haggard.
\newblock Discreteness of the volume of space from bohr-sommerfeld
  quantization.
\newblock {\em Phys. Rev. Lett.}, 107:011301, Jul 2011.

\bibitem{Bianchi:2012wb}
Eugenio Bianchi and Hal~M. Haggard.
\newblock {Bohr-Sommerfeld Quantization of Space}.
\newblock {\em Phys. Rev. D}, 86:124010, 2012.

\bibitem{PhysRevE.69.066204}
A.~D. Ribeiro, M.~A.~M. de~Aguiar, and M.~Baranger.
\newblock Semiclassical approximations based on complex trajectories.
\newblock {\em Phys. Rev. E}, 69:066204, Jun 2004.

\bibitem{PhysRevLett.79.3323}
A.~L. Xavier and M.~A.~M. de~Aguiar.
\newblock Phase-space approach to the tunnel effect: A new semiclassical
  traversal time.
\newblock {\em Phys. Rev. Lett.}, 79:3323--3326, Nov 1997.

\bibitem{MBaranger_2001}
M~Baranger, M~A~M de~Aguiar, F~Keck, H~J Korsch, and B~Schellhaaß.
\newblock Semiclassical approximations in phase space with coherent states.
\newblock {\em Journal of Physics A: Mathematical and General}, 34(36):7227,
  aug 2001.

\bibitem{10.1063/1.3583996}
Thiago~F. Viscondi and Marcus A.~M. de~Aguiar.
\newblock {Semiclassical propagator for SU(n) coherent states}.
\newblock {\em Journal of Mathematical Physics}, 52(5):052104, 05 2011.

\bibitem{10.1063/1.4936315}
Carol Braun, Feifei Li, Anupam Garg, and Michael Stone.
\newblock {The semiclassical coherent state propagator in the Weyl
  representation}.
\newblock {\em Journal of Mathematical Physics}, 56(12):122106, 12 2015.

\bibitem{Torre:2005yy}
C.~G. Torre.
\newblock {On the coherent state path integral for linear systems}.
\newblock {\em Phys. Rev. D}, 72:025004, 2005.

\bibitem{Han:2020chr}
Muxin Han and Hongguang Liu.
\newblock {Semiclassical limit of new path integral formulation from reduced
  phase space loop quantum gravity}.
\newblock {\em Phys. Rev. D}, 102(2):024083, 2020.

\bibitem{AVRAMIDI20023}
Ivan~G. Avramidi.
\newblock Heat kernel approach in quantum field theory.
\newblock {\em Nuclear Physics B - Proceedings Supplements}, 104(1):3--32,
  2002.
\newblock Proceedings of the International Meeting on Quantum Gravity and
  Spectral Geometry.

\bibitem{Dunne_2008}
Gerald~V Dunne.
\newblock Functional determinants in quantum field theory*.
\newblock {\em Journal of Physics A: Mathematical and Theoretical},
  41(30):304006, jul 2008.

\bibitem{doi:10.1142/7305}
Hagen Kleinert.
\newblock {\em Path Integrals in Quantum Mechanics, Statistics, Polymer
  Physics, and Financial Markets}.
\newblock WORLD SCIENTIFIC, 5th edition, 2009.

\bibitem{PhysRevD.18.1747}
S.~W. Hawking.
\newblock Quantum gravity and path integrals.
\newblock {\em Phys. Rev. D}, 18:1747--1753, Sep 1978.

\bibitem{doi:10.1142/1301}
G~W Gibbons and S~W Hawking.
\newblock {\em Euclidean Quantum Gravity}.
\newblock WORLD SCIENTIFIC, 1993.

\bibitem{Bodendorfer:Ha}
Norbert Bodendorfer, Thomas Thiemann, and Andreas Thurn.
\newblock New variables for classical and quantum gravity in all dimensions: I.
  \textsc{H}amiltonian analysis.
\newblock {\em Classical and Quantum Gravity}, 30(4):045001, 2013.

\bibitem{Bodendorfer:Qu}
Norbert Bodendorfer, Thomas Thiemann, and Andreas Thurn.
\newblock New variables for classical and quantum gravity in all dimensions:
  \textsc{III}. \textsc{Q}uantum theory.
\newblock {\em Classical and Quantum Gravity}, 30(4):045003, 2013.

\bibitem{Long:2020agv}
Gaoping Long and Yongge Ma.
\newblock {Polytopes in all dimensional loop quantum gravity}.
\newblock {\em Eur. Phys. J. C}, 82(41), 2022.

\bibitem{Long:2022thb}
Gaoping Long and Xiangdong Zhang.
\newblock {Gauge reduction with respect to simplicity constraint in all
  dimensional loop quantum gravity}.
\newblock {\em Phys. Rev. D}, 107(4):046022, 2023.

\bibitem{10.4310/jdg/1214459218}
Michael Kapovich and John~J. Millson.
\newblock {The symplectic geometry of polygons in Euclidean space}.
\newblock {\em Journal of Differential Geometry}, 44(3):479 -- 513, 1996.

\end{thebibliography}

\appendix
\section{Geometric interpretation of $\zeta_e$}\label{app1}

\begin{figure}[h]
 \centering
 \includegraphics[scale=0.08]{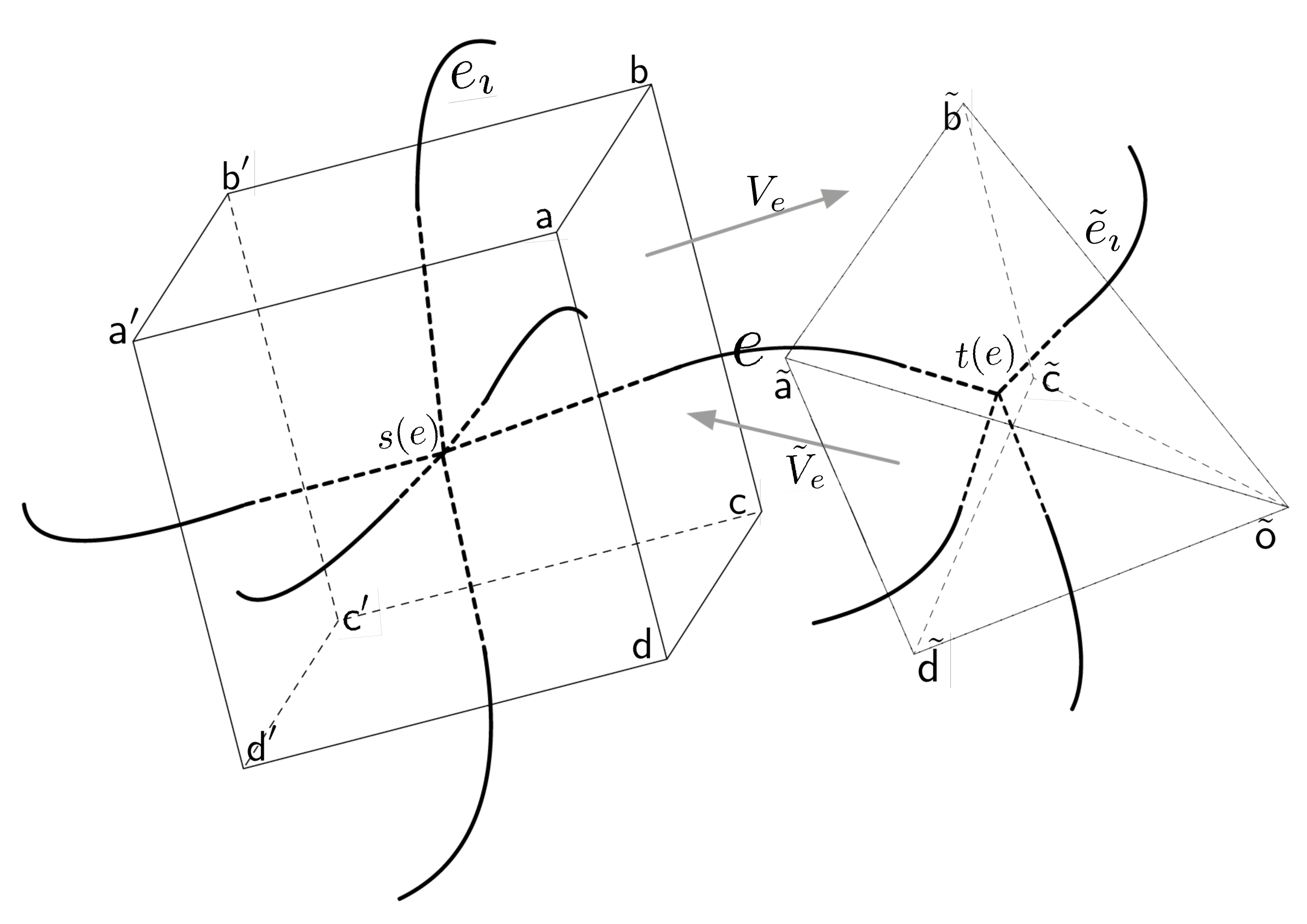}
\caption{This figure shows the geometric interpretation of each factor in the expression $h_e^{\Gamma}=n_e(V_e)e^{\zeta_e \tau_3}\tilde{n}_e(\tilde{V}_e)^{-1}$. The source point $s(e)$ of edge $e$ is dual to a hexahedron $\textsf{a}\textsf{b}\textsf{c}\textsf{d}\textsf{a}'\textsf{b}'\textsf{c}'\textsf{d}'$ and the target point $t(e)$ of edge $e$ is dual to a quadrangular pyramid $\tilde{\textsf{a}}\tilde{\textsf{b}}\tilde{\textsf{c}}\tilde{\textsf{d}}\tilde{\textsf{o}}$, while edge $e$ is dual to the two faces $\textsf{a}\textsf{b}\textsf{c}\textsf{d}$ and $\tilde{\textsf{a}}\tilde{\textsf{b}}\tilde{\textsf{c}}\tilde{\textsf{d}}$. The spin connection holonomy  $h_e^{\Gamma}=n_e(V_e)e^{\zeta_e \tau_3}\tilde{n}_e(\tilde{V}_e)^{-1}$ tells us how to glue the hexahedron $\textsf{a}\textsf{b}\textsf{c}\textsf{d}\textsf{a}'\textsf{b}'\textsf{c}'\textsf{d}'$ and quadrangular pyramid $\tilde{\textsf{a}}\tilde{\textsf{b}}\tilde{\textsf{c}}\tilde{\textsf{d}}\tilde{\textsf{o}}$ by matching the two faces $\textsf{a}\textsf{b}\textsf{c}\textsf{d}$ and $\tilde{\textsf{a}}\tilde{\textsf{b}}\tilde{\textsf{c}}\tilde{\textsf{d}}$. Specifically,  $n_e(V_e)^{-1}$ rotates the hexahedron $\textsf{a}\textsf{b}\textsf{c}\textsf{d}\textsf{a}'\textsf{b}'\textsf{c}'\textsf{d}'$ to ensure that the outward and unit normal vector $V^i_e$ of the face $\textsf{a}\textsf{b}\textsf{c}\textsf{d}$ equals to the unit vector $\delta_3^i$, while $\tilde{n}_e(\tilde{V}_e)^{-1}$ rotates the quadrangular pyramid $\tilde{\textsf{a}}\tilde{\textsf{b}}\tilde{\textsf{c}}\tilde{\textsf{d}}\tilde{\textsf{o}}$ to ensure that the outward and unit normal vector $\tilde{V}^i_e$ of the face $\tilde{\textsf{a}}\tilde{\textsf{b}}\tilde{\textsf{c}}\tilde{\textsf{d}}$ to the unit vector $-\delta_3^i$. Moreover, with $V_e^i=-\tilde{V}_e^i=\delta_3^i$, $e^{\zeta_e \tau_3}$ rotates the quadrangular pyramid $\tilde{\textsf{a}}\tilde{\textsf{b}}\tilde{\textsf{c}}\tilde{\textsf{d}}\tilde{\textsf{o}}$ to ensure that an edge of the face $\tilde{\textsf{a}}\tilde{\textsf{b}}\tilde{\textsf{c}}\tilde{\textsf{d}}$ is parallel to its corresponding edge in the face $\textsf{a}\textsf{b}\textsf{c}\textsf{d}$. The two edges from the faces $\textsf{a}\textsf{b}\textsf{c}\textsf{d}$ and $\tilde{\textsf{a}}\tilde{\textsf{b}}\tilde{\textsf{c}}\tilde{\textsf{d}}$, which will be parallel to each other with the rotation of $\tilde{\textsf{a}}\tilde{\textsf{b}}\tilde{\textsf{c}}\tilde{\textsf{d}}\tilde{\textsf{o}}$ by $e^{\zeta_e \tau_3}$, can be determined by choosing a minimal loop containing $e$. For instance, one can chooses the minimal loop $\square_e$ containing $e$, $e_\imath$ and $\tilde{e}_\imath$ , and then edges $\tilde{\textsf{a}}\tilde{\textsf{b}}$ and $\textsf{a}\textsf{b}$ dual to $\square_e$ will be parallel to each other with the rotation of $\tilde{\textsf{a}}\tilde{\textsf{b}}\tilde{\textsf{c}}\tilde{\textsf{d}}\tilde{\textsf{o}}$ by $e^{\zeta_e \tau_3}$.  }
\label{fig:labelts}
\end{figure}

{Imagine that there are two polyhedrons dual to source and target vertices $e$ where each face of the polyhedrons is dual to an edges connected to the vertices (see Fig.\ref{fig:labelts}). These two polyhedrons can be glued by the spin connection holonomy  $h_e^{\Gamma}=n_e(V_e)e^{\zeta_e\tau_3}\tilde{n}_e(\tilde{V}_e)^{-1}$ as follows. At first, $n_e(V_e)$ and $\tilde{n}_e(\tilde{V}_e)$ rotate the two polyhedrons to make $V_e=-\tilde{V}_e=\tau_3$, where $V_e$ and $\tilde V_e$ denote the normal of the 2-faces of the associated polyhedrons dual to $e$. Second, $e^{\zeta_e \tau_3}$ rotates these two polyhedrons again along the normal $V_e=-\tilde{V}_e=\tau_3$ to ensure them being aligned at the 2-faces dual to $e$. 
}
{ More explicitly, by rotating the source and target polyhedrons of $e$ to ensure $V_e=-\tilde{V}_e=\tau_3$,    $e^{\zeta_e \tau_3}$ gives the rotation of the source or target polyhedron around the unit normal vectors $V_e=-\tilde{V}_e=\tau_3$ to ensure these two polyhedrons being aligned  at the faces dual to $e$. Nevertheless, this alignment is still an undetermined notation for twisted geometry. In fact, for the Regge geometry sector of twisted geometry satisfying the shape-matching condition,  the faces dual to $e$ that are respectively from the source and target polyhedrons have identical shape, and thus the alignment is determined naturally. However, for the general twisted geometry,  the shapes of the faces dual to $e$ in  the frames of source and target polyhedrons respectively  are not necessarily identical. Thus, the strict alignment of source and target polyhedrons can not be achieved at the faces dual to $e$. Indeed, this issue has been studied in the case that both of source and target polyhedrons are tetrahedrons in Ref. \cite{PhysRevD.87.024038}, in which a relaxed alignment is proposed to construct the spin connection of the frames associated to the glued tetrahedrons  in twisted geometry, with this relaxed alignment relying on the choice of the frame on the glued triangles dual to $e$. In this article, we would like to use another type of relaxed alignment strategy adapted to arbitrary polyhedrons, in which only  one pair of edges in the glued faces are aligned as shown in Fig.\ref{fig:labelts}.  Notice that the aligned edges in the glued faces are dual to  a minimal loop $\square_e$ in $\gamma$ containing $e$, thus the definition of $\zeta_{e}$ in the holonomy of spin connection $h^{\Gamma}_{e}= n_e e^{\zeta_{e}\tau_3}\tilde{n}_e^{-1}$
depends on the choice of the minimal loop $\square_e$.}

\section{Proof of \textbf{Theorem} \ref{theorem1}}\label{app20}

\begin{proof}
    First, by definition of ${ \ell}_{v,I}, \vartheta_{v',J}$, it is easy to verify \cite{Bianchi:2010gc,10.4310/jdg/1214459218}
 \begin{equation}
 \{{ \ell}_{v,I}, \vartheta_{v',J}\}=\delta_{v,v'}\delta_{IJ}\frac{\kappa}{a^2}
\end{equation}
and
 \begin{equation}
\{{ \ell}_{v,I}, \ell_{v',J}\} =\{{ \ell}_{v,I}, \eta_e\}=\{{ \ell}_{v,I},G_{v'}\}=\{\vartheta_{v,I}, \eta_e\}= \{\vartheta_{v,I},G_{v'}\}=0,
\end{equation}
for $I,J\in\{1,2,3,4\}$, where  ${ \ell}_{v,I}:=| \vec{ \ell}_{v,I}|$.
Also, one can  verify  that  $\vartheta_{v,1}$ satisfy the following canonical Poisson brackets 
  \begin{equation}
\{\vartheta_{v,1}, \vartheta_{v,2}\}=\{\vartheta_{v,1}, \vartheta_{v,3}\}=\{\vartheta_{v,1}, \vartheta_{v,4}\}=0,
\end{equation}
by using 
 \begin{equation}
\{\vartheta_{v,1}, \mathcal{V}^i_{e_2(v)}\}\epsilon_{ijk}\mathcal{V}^j_{e_1(v)}\mathcal{V}^k_{e_2(v)}=\{\vartheta_{v,1}, \mathcal{V}^i_{e_1(v)}\}\epsilon_{ijk}\mathcal{V}^j_{e_1(v)}\mathcal{V}^k_{e_2(v)}=0,
\end{equation}
\begin{equation}
\{\vartheta_{v,1}, \mathcal{V}^i_{e_3(v)}\}\epsilon_{ijk}\mathcal{V}^j_{e_3(v)}\mathcal{V}^k_{e_4(v)}=\{\vartheta_{v,1}, \mathcal{V}^i_{e_4(v)}\}\epsilon_{ijk}\mathcal{V}^j_{e_3(v)}\mathcal{V}^k_{e_4(v)}=0,
\end{equation}
\begin{equation}
\{\vartheta_{v,1}, \ell^i_{v,1}\}\epsilon_{ijk}\ell^j_{v,1}\ell^k_{v,2}=\{\vartheta_{v,1}, \ell^i_{v,2}\}\epsilon_{ijk}\ell^j_{v,1}\ell^k_{v,2}=0,
\end{equation}
and
 \begin{equation}
\{\vartheta_{v,1}, \mathcal{V}^i_{e_5(v)}\}=\{\vartheta_{v,1}, \mathcal{V}^i_{e_6(v)}\}=\{\vartheta_{v,1}, \ell^i_{v,3}\}=\{\vartheta_{v,1}, \ell^i_{v,4}\}=0.
\end{equation}
 Following similar procedures, one can further check that 
  \begin{equation}
\{\vartheta_{v,I}, \vartheta_{v,J}\}=0
\end{equation}
for $I,J=1,2,3,4$.

Let us then check the Poisson brackets between $\varsigma_{e(v)}$ and $({ \ell}_{v,I}, \vartheta_{v',J})$. 
Notice that the definition of $\varsigma_{e(v)}$ only involves the  variables associated to the edges in $\square_e$, one has 
  \begin{equation}\label{varsigma01}
\{\varsigma_{e_1(v)},G_v\}=\{\varsigma_{e_1(v)},\eta_{e_1(v)}\mathcal{V}^i_{e_1(v)}+\eta_{e_2(v)}\mathcal{V}^i_{e_2(v)}\}=\{\varsigma_{e_1(v)}, { \ell}^i_{v,1}\}=0,
\end{equation}
  \begin{equation}\label{varsigma02}
\{\varsigma_{e_2(v)},G_v\}=\{\varsigma_{e_2(v)},\eta_{e_1(v)}\mathcal{V}^i_{e_1(v)}+\eta_{e_2(v)}\mathcal{V}^i_{e_2(v)}\}=\{\varsigma_{e_2(v)}, { \ell}^i_{v,1}\}=0,
\end{equation}
and likewise for $ \varsigma_{e_3(v)}, \varsigma_{e_4(v)}$,  $ \varsigma_{e_5(v)}, \varsigma_{e_6(v)}$, where we use $ 
G^i_v=\eta_{e_1(v)}\mathcal{V}^i_{e_1(v)}+\eta_{e_2(v)}\mathcal{V}^i_{e_2(v)}+...+\eta_{e_6(v)}\mathcal{V}^i_{e_6(v)}$. Thus we immediately have
  \begin{equation}
\{{ \ell}_{v,I},\varsigma_{e(v')}\}=\{\breve{ \ell}_{v},\varsigma_{e(v')}\}=\{\breve{ \vartheta}_{v},\varsigma_{e(v')}\}=\{\mathcal{V}_v^i,\varsigma_{e(v')}\}=0.
\end{equation}
Further, by following the geometric interpretation of $\varsigma_{e(v)}$, one has the following Poisson brackets
 \begin{equation}\label{varsigma1}
\{\varsigma_{e_1(v)}, \mathcal{V}^i_{e_2(v)}\}\epsilon_{ijk}\mathcal{V}^j_{e_1(v)}\mathcal{V}^k_{e_2(v)}=\{\varsigma_{e_1(v)}, \mathcal{V}^i_{e_2(v)}\}\delta_{ij}\mathcal{V}^j_{e_2(v)}=\{\varsigma_{e_1(v)}, \mathcal{V}^i_{e_3(v)}\}=\{\varsigma_{e_1(v)}, \mathcal{V}^i_{e_4(v)}\}=0,
\end{equation}
 \begin{equation}\label{varsigma2}
\{\varsigma_{e_2(v)}, \mathcal{V}^i_{e_1(v)}\}\epsilon_{ijk}\mathcal{V}^j_{e_1(v)}\mathcal{V}^k_{e_2(v)}=\{\varsigma_{e_2(v)}, \mathcal{V}^i_{e_1(v)}\}\delta_{ij}\mathcal{V}^j_{e_1(v)}=\{\varsigma_{e_2(v)}, \mathcal{V}^i_{e_3(v)}\}=\{\varsigma_{e_2(v)}, \mathcal{V}^i_{e_4(v)}\}=0,
\end{equation}
% \begin{equation}\label{varsigma3}\{\varsigma_{e_3(v)}, \mathcal{V}^i_{e_4(v)}\}\epsilon_{ijk}\mathcal{V}^j_{e_3(v)}\mathcal{V}^k_{e_4(v)}=\{\varsigma_{e_3(v)}, \mathcal{V}^i_{e_4(v)}\}\delta_{ij}\mathcal{V}^j_{e_4(v)}=\{\varsigma_{e_4(v)},\eta_{e_3(v)}\mathcal{V}^i_{e_3(v)}+\eta_{e_4(v)}\mathcal{V}^i_{e_4(v)}\}=0,\end{equation}
% \begin{equation}\label{varsigma4}\{\varsigma_{e_4(v)}, \mathcal{V}^i_{e_3(v)}\}\epsilon_{ijk}\mathcal{V}^j_{e_3(v)}\mathcal{V}^k_{e_4(v)}=\{\varsigma_{e_4(v)}, \mathcal{V}^i_{e_3(v)}\}\delta_{ij}\mathcal{V}^j_{e_3(v)}=\{\varsigma_{e_4(v)},\eta_{e_3(v)}\mathcal{V}^i_{e_3(v)}+\eta_{e_4(v)}\mathcal{V}^i_{e_4(v)}\}=0,\end{equation}
and likewise for  $ \varsigma_{e_3(v)}, \varsigma_{e_4(v)}$, $ \varsigma_{e_5(v)}, \varsigma_{e_6(v)}$. Combine Eqs.\eqref{varsigma01}, \eqref{varsigma02},\eqref{varsigma1} ,\eqref{varsigma2}   and Eqs.\eqref{repoi1}, one can get 
  \begin{equation}
   \{\vartheta_{v,1},\varsigma_{e_1(v)}\}=   \{\vartheta_{v,1},\varsigma_{e_2(v)}\}= 0.
\end{equation}
 Following similar procedures, one can also check that 
  \begin{equation}
   \{\vartheta_{v,I},\varsigma_{e(v')}\}= 0
\end{equation}
for $I=1,2,3,4$.

Now, we still need to calculate the Poisson bracket $ \{\varsigma_e,\varsigma_{e'}\}$. We first consider the phase space point satisfying $\mathcal{V}^i_{e'(v)}\tau_i=\tau_3$, at which one has $ \{\varsigma_{e},\zeta_{e'(v)}\}=\{\varsigma_{e},\xi_{e'(v)}\}=0$ so that $ \{\varsigma_{e},\varsigma_{e'(v)}\}=0$.   Also, it is easy to verify that $ \{\varsigma_e,\varsigma_{e'}\}$ is a constant by using the Jacobi identity
\begin{equation}
\{ \cdot,\{\varsigma_e,\varsigma_{e'}\}\}=-\{ \varsigma_e,\{\varsigma_{e'},\cdot\}\}-\{ \varsigma_{e'},\{\cdot,\varsigma_{e}\}\}=0,
\end{equation}
where $\cdot$ is an arbitrary elements in the set $\{\eta_e,{ \ell}_{v,I}, \vartheta_{v,I}, \breve{\ell}_v,\breve{\vartheta}_{v}, \mathcal{V}_v^i\}$.  Then, one can conclude that 
\begin{equation}
\{\varsigma_e,\varsigma_{e'}\}=0
\end{equation}
on the whole twisted geometry space $P_\gamma^+$. This finish the proof. 
\end{proof}

\end{document}